\documentclass{article}

\usepackage{arxiv}

\usepackage[utf8]{inputenc} 
\usepackage[T1]{fontenc}    
\usepackage{hyperref}       
\usepackage{url}            
\usepackage{booktabs}       
\usepackage{amsfonts}       
\usepackage{nicefrac}       
\usepackage{microtype}      
\usepackage{lipsum}		
\usepackage{graphicx}
\usepackage{natbib}
\usepackage{doi}

\usepackage{amsmath}
\usepackage{amssymb}
\usepackage{appendix}
\usepackage{float}
\usepackage{multirow}
\usepackage{xcolor}
\definecolor{ARR}{rgb}{1,0,0} 
\definecolor{ARR2}{rgb}{1,0,0}
\definecolor{ARR3}{rgb}{1,0,0}

\title{A semi-analytical solution for the lubrication force between two spheres approaching in viscoelastic fluids described by the Oldroyd-B model under small Deborah numbers}

\author{ Alan Rosales-Romero\\
	Facultad de Química, Departamento de Ingeniería Química, Universidad Nacional Autónoma de México (UNAM),\\ Ciudad Universitaria, Coyoacán, Ciudad de México, 04510, Mexico.\\
    Basque Center for Applied Mathematics (BCAM), Alameda de Mazarredo 14, Bilbao, 48009, Spain.\\
	\And
	Adolfo Vázquez-Quesada\\
	Departamento de Física Fundamental, UNED, Apartado 60141, Madrid, 28080, Spain.\\
	\AND
	Marco Ellero \\
	Basque Center for Applied Mathematics (BCAM),
	Alameda de Mazarredo 14, Bilbao, 48009, Spain.\\
        IKERBASQUE, Basque Foundation for Science, Calle de María Díaz de Haro 3, Bilbao, 48013, Spain.\\
        Complex Fluids Research Group, Swansea University, Bay Campus, Swansea, SA1 8EN, United Kingdom.\\
	\And
	J. Esteban López-Aguilar\thanks{Corresponding autohr: jelopezaguilar@quimica.unam.mx} \\
	Facultad de Química, Departamento de Ingeniería Química, Universidad Nacional Autónoma de México (UNAM),\\ Ciudad Universitaria, Coyoacán, Ciudad de México, 04510, Mexico.\\
}

\renewcommand{\shorttitle}{A semi-analytical solution for the lubrication force between two spheres approaching in viscoelastic fluids described by the Oldroyd-B model under small Deborah numbers}

\hypersetup{
pdftitle={A semi-analytical solution for the lubrication force between two spheres approaching in viscoelastic fluids described by the Oldroyd-B model under small Deborah numbers},
pdfauthor={A. Rosales-Romero, A. Vázquez-Quesada, M. Ellero and J.E. López-Aguilar},
pdfkeywords={Semi-analytical solution, normal lubrication, Oldroyd-B model, viscoelasticity, squeezing flow of two spheres, normal-stress differences.},
}

\begin{document}
\maketitle

\begin{abstract}
Viscoelastic fluids play a critical role in various engineering and biological applications, where their lubrication properties are strongly influenced by relaxation times ranging from microseconds to minutes. Although the lubrication mechanism for Newtonian fluids is well-established, its extension into viscoelastic materials\textemdash particularly under squeezing flow conditions\textemdash remains less explored. This study presents a semi-analytical solution for the lubrication force between two spheres approaching in a Boger fluid under small Deborah numbers. Unlike previous works that assumed a Newtonian velocity field, we derive the velocity profile directly from the mass-momentum conservation and Oldroyd-B constitutive equations using lubrication theory and order-of-magnitude analysis techniques.
Under steady-state conditions, viscoelasticity induces a marginal increase in the surface-to-surface normal force as a result of the increased pressure required to overcome the resistance originating from the first normal-stress difference. Transient analyses reveal that the normal lubrication force is bounded by two Newtonian plateaus and is non-symmetric as the spheres approach or separate. Our findings highlight the role of viscoelasticity in improving load capacity and provide new insights for modelling dense particle suspensions in Boger fluids, where short-range interactions dominate.
\end{abstract}


\section{Introduction}
Viscoelastic fluids are fundamental in various technological and biological contexts involving close-contact flows, such as lubrication of sliding surfaces, synthetic transport in living organisms, and synovial joint dynamics \citep{AhmedBiancofioreJNNFM2021, UrryACIEE1993, AteshianJB2009}. Their defining feature\textemdash a wide range of relaxation times\textemdash leads to complex non-linear responses governed by normal-stress differences \citep{ZhangLiJEM2005, NametalJCA2015}. Despite extensive studies on Newtonian suspensions \citep{HoffmanJRheol1972,ZarragaJRheol2000,SciroccoetalJR2005,GuazzelliPouliquenJFM2018}, suspensions in viscoelastic media remain less explored. Experiments with viscoelastic fluids, including Boger fluids characterised by a nearly constant shear viscosity and pronounced elasticity \citep{OldroydPRS1950, BogerJNNFM1977}, reveal distinctive behaviours, such as early shear thickening, complex stress scaling, and matrix–microstructure interactions \citep{ZarragaJRheol2001, SciroccoetalJNNFM2004, DaiTannerRA2020}.
Suspensions in viscoelastic fluids also exhibit particle migration, chaining, and flow-induced structuring \citep{MicheleetalRA1997, PasquinoetalJCIS2013, GiudiceetalLC2013}. Microscopy studies in shear-thinning media show string-like particle alignment at low Weissenberg numbers, governed by fluid rheology and particle size \citep{PasquinoetalRA2010}. Settling experiments indicate that viscoelasticity and shear-thinning enhance migration, unlike Newtonian fluids, for which such effects are suppressed \citep{SciroccoetalJR2005,DAvinoalRA2012}. In confined microfluidic environments, particles can self-organise into ordered structures due to coupled inertial and viscoelastic forces, underscoring the significance of confinement and non-Newtonian flow features in particle-laden systems \citep{VilloneEtalCF2011, GiudiceAvinoJPM2025}.

Numerical studies on viscoelastic suspensions have largely focused on dilute to semi-dilute regimes (volume fractions up to $\phi = 0.3$), offering valuable insights into particle-laden viscoelastic flows through a variety of computational frameworks.
\cite{VazquezQEtalJFM2019} employed Smoothed Particle Hydrodynamics (SPH), incorporating a GENERIC-based viscoelastic model, to simulate rigid particle suspensions. They observed a plateau in relative viscosity at low Deborah numbers (\textit{De}), followed by shear thickening at higher \textit{De} due to extensional features in the matrix. \cite{JainShaqfehJR2021} used body-fitted (BF) and immersed-boundary (IB) methods to study Boger fluid suspensions, finding increases in per-particle viscosity and the first normal-stress coefficient with strain, particle volume fraction, and Weissenberg number. Interestingly, fluid stress remained independent of $\phi$, while the stresslet rose due to enhanced hydrodynamic interactions. \cite{VilloneetalJR2021} investigated immersed-boundary-method numerical simulations of small-amplitude oscillatory shear in shear-thinning and viscoelastic matrices, showing inertia-induced elasticity and Carreau number–dependent shear-thinning effects. Viscoelastic moduli varied with \textit{De} and viscosity ratio, but not with the Giesekus mobility factor.
Several methodologies have expanded into diverse modelling tool sets: \cite{KuronetalEPJE2021} proposed a lattice Boltzmann method (LBM) for Oldroyd-B fluids, preserving stress and energy in complex flows. \cite{MatsuokaetalJFM2021} and \cite{YamamotoetalSM2021} applied the smoothed profile method (SPM), effectively predicting shear thickening and accommodating complex particulate systems. \cite{ZhangShaqfehJR2023} combined 3D finite-volume simulations with experiments to study shear-thinning suspensions, linking reduced viscosity to diminished stresslet contributions and particle-induced stresses. Recent extensions of these frameworks include the lattice Boltzmann–smoothed profile hybrid method (LBM-SPM) by \cite{LeeKARJ2023}, and the thermally augmented SPM by \cite{NakayamaetalJCP2024}, which enable accurate micro-rheological analysis in thermally fluctuating viscoelastic fluids. Previous simulations of passive microrheology for a colloidal particle immersed in a thermodynamically-consistent fluctuating Oldroyd-B fluid were already presented earlier by \cite{VazquezQuesadaEtalMN2012}.

Despite the significant amount of numerical work performed in the dilute and semi-dilute regimes, examining dense suspensions where $\phi > 0.3$ has been much less investigated. For suspensions with Newtonian matrices, recent advances in simulations of the rheology of concentrated suspensions—particularly dense, non-colloidal systems—highlight the crucial role of microstructure, contact mechanics, and interparticle forces in governing non-Newtonian flow phenomena \citep[see][]{MorrisRA2009,SetoMorrisDennPRL2013,WyartCatesPRL2014,TannerPF2018}. To simulate suspensions with viscoelastic matrices at high volume fraction, however, it is necessary to specify how particles hydrodynamically interact when they are in close proximity, i.e., proposing a model for interparticle lubrication as done for Newtonian fluids in \cite{PrasannaVQElleroJCP2021}, for example. Currently, there are no numerical techniques available for analysing dense viscoelastic suspensions since there is a lack of generalisations regarding non-Newtonian lubrication forces acting between particles. Novel lubrication force models would facilitate the development of precise and efficient reduced-order methods of the Lubrication Dynamics (LD) or Discrete Element Method (DEM) type \citep{PrasannaVQElleroJCP2021,RuizLopezetalJRheol2023tribo,RosalesetalJNNFM2024}.

Analytical investigations into the squeezing flow between nearly-touching spheres have provided valuable insights into lubrication interactions across different rheological regimes. In the Newtonian case, \cite{ONeillASR1969} presented the exact solutions for the asymmetrical slow viscous flows of an infinite fluid caused by either the rotation or by the translation of spheres along directions perpendicular to their line of centres, and \cite{JeffreyMATH1982} employed an asymptotic analysis to evaluate the force between two spheres of unequal radii approaching each other with equal and opposite velocities under Stokes conditions. Extending this framework to non-Newtonian settings, \cite{RodinJNNM1996} derived asymptotic solutions for Power-Law fluids, showing that resistance is dominated by the gap flow only when the Power-Law index is $m \geq 1/3$, while for $m < 1/3$, outer-region contributions become significant. Further recent advancement has been achieved by employing both a bi-viscous shear thinning in \cite{VazquezElleroPF2016}, and a shear-thickening fluid model in \cite{VazquezetalPF2018}, to obtain approximate expressions for the generalised lubrication force between adjacent spheres in both kinds of fluids.

In contrast to inelastic lubricating fluids, studies focusing on viscoelastic media are comparatively limited. \cite{XuetalAMM2004} used a perturbation method based on the Reynolds’ lubrication theory to estimate forces between rigid spheres in a second-order fluid, while \cite{DandekarArdekaniPF2021} employed a perturbation expansion in the Deborah number to derive viscoelastic contributions to hydrodynamic forces, including lift components along the line connecting sphere centres. Recent work by \cite{ZhengetalPF2023,ZhengetalPF2024} applied a lubrication approach and an order-of-magnitude analysis to squeezing flows in upper-convected Maxwell fluids, offering estimates for pressure, shear and normal stresses. \cite{RuangStonePRF2024} also examined Oldroyd-B fluids and showed that viscoelastic effects can either enhance or reduce resistance depending on the order of approximation and motion direction. Despite these advances, most studies  assume that the velocity field remains parabolic and quasi-steady, as in Newtonian \citep{PhanTannerJNNM1984,PhanetalJNNM1985,ZhengetalPF2023} or shear-thinning fluids \citep{McClellandFinlaysonJNNFM1983, BairTT2015}, thereby neglecting direct modifications to flow kinematics arising from viscoelastic normal-stress differences. This simplification, while practical, may limit model accuracy in regimes where elastic effects significantly influence flow structure.

In this study, we present a semi-analytical solution for the normal lubrication force between two rigid spheres approaching each other in a viscoelastic fluid, modelled using the Oldroyd-B \citep{OldroydPRS1950} constitutive equation under small Deborah numbers. Unlike previous approaches, which assume Newtonian velocity fields, our formulation directly solves the fully-coupled mass, momentum, and constitutive equations using lubrication theory and order-of-magnitude analysis. We consider both steady and transient regimes under constant approach or departure velocities without prescribing a velocity profile a priori. Our findings enable future numerical LD/DEM simulations of dense particle suspensions within Boger-type viscoelastic fluids. The structure of this paper is the following: in Section \ref{Sec:Equations}, the flow settings, and the conservation and constitutive equations are described for the approaching-spheres flow system under consideration. In Section \ref{Sec:Steady}, the steady-state problem is discussed, whilst Section \ref{Sec:Transient} focuses on the transient state. In Section \ref{Sec:ModelPre}, predictions with a real Boger fluid characterised with the Oldroyd-B model are reported. Finally, conclusions are offered in Section \ref{Sec:Conclusions}.

\section{Flow setting, conservation and constitutive equations} \label{Sec:Equations}

\subsection{Squeezing flow between two spherical particles}

\begin{figure}
\centering
\includegraphics[width=0.35\linewidth]{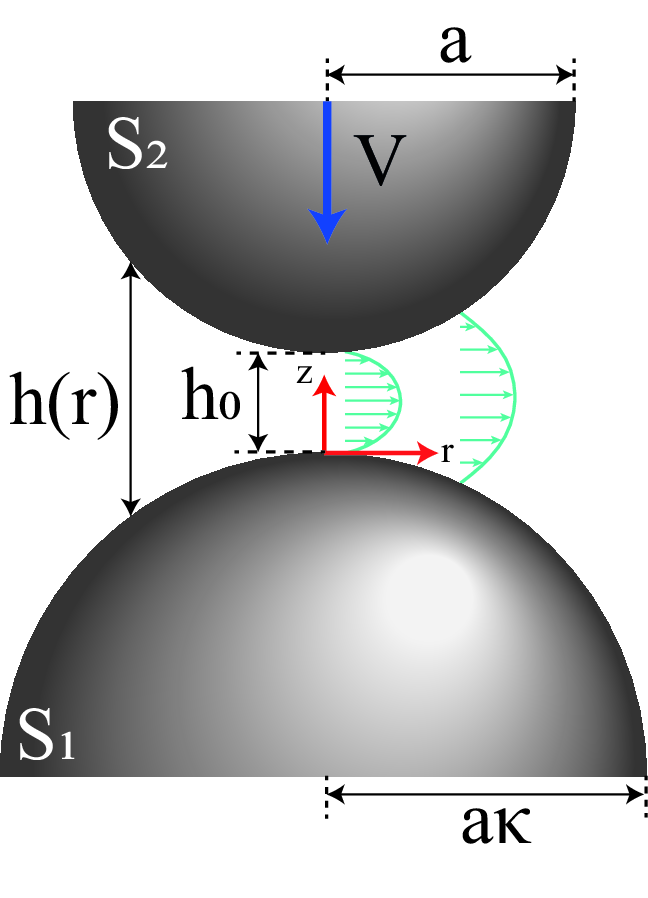}
\caption{Schematics of two spheres approaching with relative velocity $V$. The minimum distance between the sphere contours is $h_0$.}
\label{Fig:twoparticles}
\end{figure}

The flow settings are analogous to that reported by \cite{VazquezElleroPF2016}, where the approaching motion of two spheres with a fluid being squeezed in between their surfaces is considered. In this work, we start with a viscoelastic fluid between two spheres, $S_1$ and $S_2$, that are smooth and rigid, with radii $a \kappa$ and $a$, respectively, and $\kappa \geq 1$, as illustrated in Figure \ref{Fig:twoparticles}. Here, the particle on the bottom ($S_1$) is kept static and the top one ($S_2$) approaches with constant velocity $V$. In the limit of narrow sphere separations, the particle contours can be approximated by paraboloids, where the lower and upper sphere contours are described as follows: $z_1(r) = -\frac{r^2}{2a \kappa}$ and $z_2(r) = h_0 + \frac{r^2}{2a}$, respectively. Therefore, the distance between their surfaces is $h(r) = z_2(r) - z_1(r) = h_0 \left( 1 + \frac{r^2}{2ah_0 \kappa_a} \right)$, where $h_0$ is the minimum gap width between spheres (at $r=0$) and $\frac{1}{\kappa_a}=1+\frac{1}{\kappa}$. The reference cylindrical coordinate system is chosen as $\{ r, \theta, z \}$, and considering an axisymmetric case with irrotational motion, i.e., $v_{\theta}=0$ and $\frac{\partial (\cdot) }{\partial \theta} = 0$, the velocity field is $\boldsymbol{v}= \left( v_r(t, r, z), \; 0, \; v_z(t, r, z) \right)$, where an explicit temporal and spatial dependence may be specified.

\subsection{Conservation equations and constitutive equation}

The mass and momentum conservation equations for an incompressible fluid ignoring the body force are:

\begin{equation} 
\boldsymbol{\nabla} \cdot \boldsymbol{v} = 0,
\end{equation}
\
\begin{equation} 
\rho \frac{D \boldsymbol{v}}{D t} = \boldsymbol{\nabla} \cdot \boldsymbol{\Pi}.
\end{equation}
\
Here, the total stress is:

\begin{equation}
\boldsymbol{\Pi} = -p\boldsymbol{I} + 2 \eta_{s} \boldsymbol{D} + \boldsymbol{\tau},
\label{Eq:TotalStress}
\end{equation}
\
where $p$ is the isotropic pressure, $\boldsymbol{I}$ is the identity tensor, and the remaining two terms compose the extra-stress. The extra-stress can be decomposed into two parts: a purely-viscous component, $2 \eta_{s} \boldsymbol{D}$, that corresponds to the Newtonian solvent contribution to the total stress ($\eta_s$ is the solvent viscosity and $\boldsymbol{D}$ is the rate-of-deformation tensor, defined as $\boldsymbol{D} = \frac{1}{2} \left( \nabla \boldsymbol{v} + (\nabla \boldsymbol{v})^T \right)$), and the viscoelastic-stress $\boldsymbol{\tau}$ whose evolution is governed by the upper-convected Maxwell model, as follows:

\begin{equation}
\boldsymbol{\tau} + \lambda_1 \overset{\bigtriangledown}{\boldsymbol{\tau}} =  2 \eta_{p} \boldsymbol{D}.
\label{Eq:OB1} 
\end{equation}
\
Here, $\lambda_1$ is the relaxation time, $\eta_p$ is the polymeric viscosity, and $\overset{\bigtriangledown}{\boldsymbol{\tau}}$ denotes the upper-convected time derivative, given by:
\begin{equation}
\overset{\bigtriangledown}{\boldsymbol{\tau}} = \frac{\partial \boldsymbol{\tau}}{\partial t} + \boldsymbol{v} \cdot \boldsymbol{\nabla} \boldsymbol{\tau} - \boldsymbol{\nabla} \boldsymbol{v}^T \cdot \boldsymbol{\tau} - \boldsymbol{\tau} \cdot \boldsymbol{\nabla} \boldsymbol{v}.
\label{Eq:UCD}
\end{equation}
\
Additional parameters are introduced for viscosity ratios: the total viscosity of the viscoelastic solution is $\eta_0 = \eta_s + \eta_p$. A solvent-fraction measure $\beta = \frac{\eta_s}{\eta_0}$ may be defined to account for the relative strength of the solute-to-solvent contributions. From here, one may recover the solvent viscosity $\eta_s = \eta_0 \beta$ and the polymeric viscosity $\eta_p = \eta_0 ( 1 - \beta )$.

\subsection{Problem statement}

For an axisymmetric deformation in cylindrical coordinates, the continuity equation takes the following form:
\begin{equation} 
\frac{1}{r} \frac{\partial \left( rv_{r} \right)}{\partial r} + \frac{\partial v_{z}}{\partial z} = 0,
\label{Eq:Cont}
\end{equation} 
\
the momentum balance equation in $r-$direction becomes:
\begin{equation}
\rho \left( \frac{\partial v_{r}}{\partial t} + v_{r} \frac{\partial v_{r}}{\partial r} + v_{z} \frac{\partial v_{r}}{\partial z} \right) = -\frac{\partial p}{\partial r} + \eta_{s} \left[ \frac{\partial}{\partial r}\left( \frac{1}{r} \frac{\partial (r v_r)}{\partial r}\right) + \frac{\partial^2 v_r}{\partial z^2} \right] + \\
\left( \frac{1}{r} \frac{\partial \left(r\tau_{rr} \right)}{\partial r} + \frac{\partial \tau_{rz}}{\partial z} - \frac{\tau_{\theta \theta}}{r} \right),
\label{Eq:Mom_r}
\end{equation}
\
and in $z-$direction:
\begin{equation} 
\rho \left( \frac{\partial v_{z}}{\partial t} + v_{r} \frac{\partial v_{z}}{\partial r} + v_{z} \frac{\partial v_{z}}{\partial z} \right) = -\frac{\partial p}{\partial z} + \eta_{s} \left[ \frac{1}{r} \frac{\partial}{\partial r} \left( r \frac{\partial v_z}{\partial r} \right) + \frac{\partial^2 v_z}{\partial z^2} \right] + \\
\left( \frac{1}{r} \frac{\partial \left(r\tau_{rz} \right)}{\partial r} + \frac{\partial \tau_{zz}}{\partial z} \right).
\label{Eq:Mom_z}
\end{equation}  
\
Given the symmetry condition in the $\theta-$direction, the associated stress equations are inherently satisfied. To close the problem, we impose no-slip boundary conditions over the surface of the spheres, as follows:

\begin{equation}
\begin{split}
v_r & = 0, \quad v_z = 0 \quad \text{on } z = z_1(r), \\
v_r & = 0, \quad v_z = -V \quad \text{on } z = z_2(r).
\end{split}
\label{Eq:CF}
\end{equation}
\\
With the velocity field $\boldsymbol{v}= \left( v_r(t, r, z), \; 0, \; v_z(t, r, z) \right)$, the rate-of-deformation tensor becomes:

\begin{equation}
\boldsymbol{D} = \frac{1}{2}
\begin{pmatrix}
2 \frac{\partial v_{r}}{\partial r} & 0 & \frac{\partial v_{r}}{\partial z} + \frac{\partial v_{z}}{\partial r} \\
0 & 2 \frac{v_r}{r} & 0\\
\frac{\partial v_{r}}{\partial z} + \frac{\partial v_{z}}{\partial r} & 0 & 2 \frac{\partial v_{z}}{\partial z}
\end{pmatrix}.
\label{Eq:D}
\end{equation} 

Substituting (\ref{Eq:UCD}) and (\ref{Eq:D}) in (\ref{Eq:OB1}), the viscoelastic-stress components emerge as follows:

\begin{equation}
\tau_{rr} + \lambda_1 \left( \frac{\partial \tau_{rr}}{\partial t} + v_r \frac{\partial \tau_{rr}}{\partial r} + v_z \frac{\partial \tau_{rr}}{\partial z} - 2 \left( \tau_{rr} \frac{\partial v_r}{\partial r} + \tau_{zr} \frac{\partial v_r}{\partial z} \right)  \right) = 
2 \eta_0 (1 - \beta) \frac{\partial v_r}{\partial r}, 	
\label{Eq:Tprr}
\end{equation} \
\begin{multline}
\tau_{rz} + \lambda_1 \left( \frac{\partial \tau_{rz}}{\partial t} + v_r \frac{\partial \tau_{rz}}{\partial r} + v_z \frac{\partial \tau_{rz}}{\partial z} - \tau_{rz} \left( \frac{\partial v_r}{\partial r} + \frac{\partial v_z}{\partial z} \right) - \tau_{zz} \frac{\partial v_r}{\partial z} - \tau_{rr} \frac{\partial v_z}{\partial r} \right) = \\
\eta_0 (1 - \beta) \left( \frac{\partial v_z}{\partial r} + \frac{\partial v_r}{\partial z} \right), 
\label{Eq:Tprz}
\end{multline} \
\begin{equation}
\tau_{\theta\theta} + \lambda_1 \left( \frac{\partial \tau_{\theta\theta}}{\partial t} + v_r \frac{\partial \tau_{\theta\theta}}{\partial r} + v_z \frac{\partial \tau_{\theta\theta}}{\partial z} - 2 \tau_{\theta \theta} \frac{v_r}{r} \right) = 2 \eta_0 (1 - \beta) \frac{v_r}{r},
\label{Eq:Tptt}
\end{equation} \
\begin{equation}
\tau_{zz} + \lambda_1 \left( \frac{\partial \tau_{zz}}{\partial t} + v_r \frac{\partial \tau_{zz}}{\partial r} + v_z \frac{\partial \tau_{zz}}{\partial z}  - 2 \left( \tau_{rz} \frac{\partial v_z}{\partial r} + \tau_{zz} \frac{\partial v_z}{\partial z} \right) \right) = \\
2 \eta_0 (1 - \beta) \frac{\partial v_z}{\partial z}.
\label{Eq:Tpzz}
\end{equation}

\subsection{Dimensionless equations}

The system of partial differential equations to solve is given by (\ref{Eq:Cont})-(\ref{Eq:Mom_z}), and (\ref{Eq:Tprr})-(\ref{Eq:Tpzz}).
Under the assumption of very small interparticle gaps, one can assume that the ratio between the gap and the reduced radius is small, i.e., $\epsilon = \frac{h_0}{a} \ll 1$; this allows to adopt an order-of-magnitude analysis of the system of partial differential equations through lubrication theory \citep{KimKarrila1997}.
With this aim, we start by non-dimensionalising the system above in non-dimensional stretched coordinates $\{ R, Z \}$ \citep{JeffreyMATH1982, RodinJNNM1996}:

\begin{equation}
\begin{split}
R & = \frac{r}{a \epsilon^{1/2}}, \; Z = \frac{z}{h_0}, \; T = \frac{V t}{h_0}, \; u = \frac{v_r}{U}, \; w = \frac{v_z}{V}, \; P =\frac{ \epsilon^{1/2} p}{\eta_0 U / h_0}, \\
H(R) & = \frac{h(r)}{h_0} = 1 + \frac{R^2}{2 \kappa_a}, \; \boldsymbol{S} = \frac{\boldsymbol{\tau}}{\eta_0 U / h_0}, \; De = \frac{U \lambda_1}{h_0}, \; Re = \frac{\rho U h_0}{\eta_0},
\end{split}	
\label{Eq:adimvar}
\end{equation}
\
where the approaching velocity $V$ is the characteristic velocity in the $z-$direction, and $U = \frac{V}{\epsilon^{1/2}}$ is the characteristic velocity in the $r-$direction (the direction of the flow process), derived from (\ref{Eq:Cont}). Here, the dimensionless group-numbers of Deborah ($De$) and Reynolds ($Re$) measure the ratio between elastic and viscous forces, and the ratio between inertial and viscous forces, respectively. Moreover, since the \textit{Re} number between particles is small, temporal and convective terms may be ignored in the momentum balance. Hence, the conservation equations (\ref{Eq:Cont})-(\ref{Eq:Mom_z}) appear in dimensionless form, as follows:

\begin{equation} 
\frac{1}{R} \frac{\partial \left( R u \right)}{\partial R} + \frac{\partial w}{\partial Z} = 0, 
\label{Eq:Cont_Adim}
\end{equation} \
\begin{equation} 
- \frac{\partial P}{\partial R} + \beta \left[ \epsilon \frac{\partial}{\partial R} \left( \frac{1}{R} \frac{\partial (R u)}{\partial R} \right) + \frac{\partial ^2 u}{\partial Z ^2} \right] + \frac{\partial S_{rz}}{\partial Z} + \epsilon^{1/2} \frac{1}{R} \frac{\partial \left(R S_{rr}\right)}{\partial R} \\
- \epsilon^{1/2} \frac{S_{\theta \theta}}{R} = 0,
\label{Eq:Mom_rAdim}
\end{equation} \
\begin{equation} 
- \frac{\partial P}{\partial Z} + \beta \left[ \epsilon^2 \frac{1}{R} \frac{\partial}{\partial R} \left( R \frac{\partial w}{\partial R} \right) + \epsilon \frac{\partial ^2 w}{\partial Z ^2} \right] + \epsilon \frac{1}{R} \frac{\partial \left(R S_{rz}\right)}{\partial R} + \epsilon^{1/2} \frac{\partial S_{zz}}{\partial Z} = 0.
\label{Eq:Mom_zAdim}
\end{equation} 
\\
To show explicitly the effects of the viscoelasticity through normal-stress differences that may arise, a re-casted expression for the momentum balance equation in the \textit{r}-direction can be obtained by adding and subtracting $\frac{S_{zz}}{R}$ on (\ref{Eq:Mom_rAdim}), which then may be rearranged as:
\begin{equation}
- \frac{\partial P}{\partial R} + \beta \left[ \epsilon \frac{\partial}{\partial R} \left( \frac{1}{R} \frac{\partial (R u)}{\partial R} \right) + \frac{\partial ^2 u}{\partial Z ^2} \right] + \frac{\partial S_{zr}}{\partial Z} + \\
\epsilon^{1/2} \left( \frac{\partial S_{rr}}{\partial R} + \frac{N_{1}}{R} + \frac{N_{2}}{R} \right) = 0.
\label{Eq:Mom_rAdim_N}
\end{equation}
\\
Here, the first and second normal-stress differences are defined as $N_1 = S_{rr} - S_{zz}$ and $N_2 = S_{zz} - S_{\theta \theta}$, respectively.
Under the same rationale, the viscoelastic-stress components in (\ref{Eq:Tprr})-(\ref{Eq:Tpzz}) become:

\begin{equation}
S_{rr} + De \left( \frac{\partial S_{rr}}{\partial T} + u \frac{\partial S_{rr}}{\partial R} + w \frac{\partial S_{rr}}{\partial Z} - 2 \left( S_{rr} \frac{\partial u}{\partial R} + \frac{S_{zr}}{\epsilon^{1/2}} \frac{\partial u}{\partial Z} \right)  \right) = \\
2 (1 - \beta) \epsilon^{1/2} \frac{\partial u}{\partial R},
\label{Eq:TprrAdim}
\end{equation} \\
\begin{multline}
S_{rz} + De \left( \frac{\partial S_{rz}}{\partial T} + u \frac{\partial S_{rz}}{\partial R} + w \frac{\partial S_{rz}}{\partial Z} - S_{rz} \left( \frac{\partial u}{\partial R} + \frac{\partial w}{\partial Z} \right) - \frac{S_{zz}}{\epsilon^{1/2}} \frac{\partial u}{\partial Z} - \epsilon^{1/2} S_{rr} \frac{\partial w}{\partial R} \right) = \\
 (1 - \beta) \left( \epsilon \frac{\partial w}{\partial R} + \frac{\partial u}{\partial Z} \right),
\label{Eq:TprzAdim}
\end{multline} \\
\begin{equation}
S_{\theta\theta} + De \left( \frac{\partial S_{\theta\theta}}{\partial T} + u \frac{\partial S_{\theta\theta}}{\partial R} + w \frac{\partial S_{\theta\theta}}{\partial Z} - 2S_{\theta \theta \frac{u}{R}} \right) = 2 (1 - \beta) \epsilon^{1/2} \frac{u}{R},
\label{Eq:TpttAdim}
\end{equation} \\
\begin{equation}
S_{zz} + De \left( \frac{\partial S_{zz}}{\partial T} + u \frac{\partial S_{zz}}{\partial R} + w \frac{\partial S_{zz}}{\partial Z}  - 2 \left( \epsilon^{1/2} S_{rz} \frac{\partial w}{\partial R} + S_{zz} \frac{\partial w}{\partial Z} \right) \right) = \\
2 (1 - \beta) \epsilon^{1/2} \frac{\partial w}{\partial Z}.
\label{Eq:TpzzAdim}
\end{equation} 

\section{Steady-state case} \label{Sec:Steady}

Under the assumption of spheres being very close between each other, i.e., $\epsilon = \frac{h_0}{a} \ll 1$, the terms in (\ref{Eq:TprrAdim})-(\ref{Eq:TpzzAdim}) with order-of-magnitude $O(\epsilon)^{1/2}$ and higher may be considered negligible. In this first section, we also consider a steady-state approximation, making null the temporal partial-derivative terms, i.e., $\frac{\partial (\cdot)}{\partial T} = 0$ in the viscoelastic stress. Moreover, to get a first approximation, we consider a weakly-viscoelastic instance through a small \textit{De}-number limiting case, considering \textit{De}-numbers with a similar magnitude to $\epsilon^{1/2}$, such that $\frac{De}{\epsilon^{1/2}} \sim O(1)$. Under such set of assumptions, the viscoelastic-stress components reduce to:

\begin{equation}
S_{rr} - 2 \left( \frac{De}{\epsilon^{1/2}} \right) S_{zr} \frac{\partial u}{\partial Z} = 0,
\label{Eq:TprrAdim_s}
\end{equation} \
\begin{equation}
S_{rz} - \left( \frac{De}{\epsilon^{1/2}} \right) S_{zz} \frac{\partial u}{\partial Z} = (1 - \beta) \frac{\partial u}{\partial Z},
\label{Eq:TprzAdim_s}
\end{equation} \
\begin{equation}
S_{\theta\theta} = 0,
\label{Eq:TpttAdim_s}
\end{equation} \
\begin{equation}
S_{zz} = 0.
\label{Eq:TpzzAdim_s}
\end{equation} 

In this case, $S_{zz} = S_{\theta \theta} = S_{r \theta} = S_{\theta z} = 0$. Hence, only two components of the viscoelastic-stress are different to zero:

\begin{equation}
S_{rz} = (1 - \beta) \frac{\partial u}{\partial Z},
\label{Eq:TprrAdim_sol}
\end{equation} \
\begin{equation}
S_{rr} = 2 (1 - \beta) \frac{De}{{\epsilon^{1/2}}} \left( \frac{\partial u}{\partial Z} \right)^2,
\label{Eq:TprzAdim_sol}
\end{equation}
\\
and, therefore, $N_1 = S_{rr}$ and $N_2 = 0$. The first normal-stress difference is led by the normal stress in the flow direction, i.e., $S_{rr}$. Conversely, the second normal-stress difference is zero and it will have no influence in the motion of the fluid.
Then, the momentum balance equations are obtained by substituting (\ref{Eq:TprrAdim_sol}) and (\ref{Eq:TprzAdim_sol}) in (\ref{Eq:Mom_zAdim}) and (\ref{Eq:Mom_rAdim_N}), respectively, such that:

\begin{equation} 
- \frac{\partial P}{\partial R}+ \frac{\partial ^2 u}{\partial Z ^2} + \frac{2(1 - \beta)De}{R} \left( \frac{\partial u}{\partial Z} \right) ^2 = 0,
\label{Eq:Mom_rAdim_s}
\end{equation} \
\begin{equation} 
\frac{\partial P}{\partial Z} = 0,
\label{Eq:Mom_zAdim_s}
\end{equation} 

From examination of (\ref{Eq:Mom_rAdim_s}) and (\ref{Eq:Mom_zAdim_s}), one may conclude that  the pressure $P$-dependence appears only in the radial direction. In (\ref{Eq:Mom_rAdim_s}), the viscous response is given by the shear-stress, such  that $\frac{\partial ^2 u}{\partial Z ^2}$ manifests the viscous resistance to the rate-of-deformation. Complementarily, the viscoelastic response is led by the first  normal-stress difference, such that $\frac{2(1 - \beta)De}{R} \left( \frac{\partial u}{\partial Z} \right) ^2$, represents a measure of the viscoelastic resistance of deformation. This viscoelastic term in (\ref{Eq:Mom_rAdim_s}) represents an additional resistance to shearing deformation that the fluid exhibits in the radial direction, that results from the stretching and alignment of the material along the streamlines \citep{Byronetal1BOOK1987}, and which the radial pressure gradient must overcome.

\begin{figure}
\centering
\includegraphics[width=0.6\linewidth]{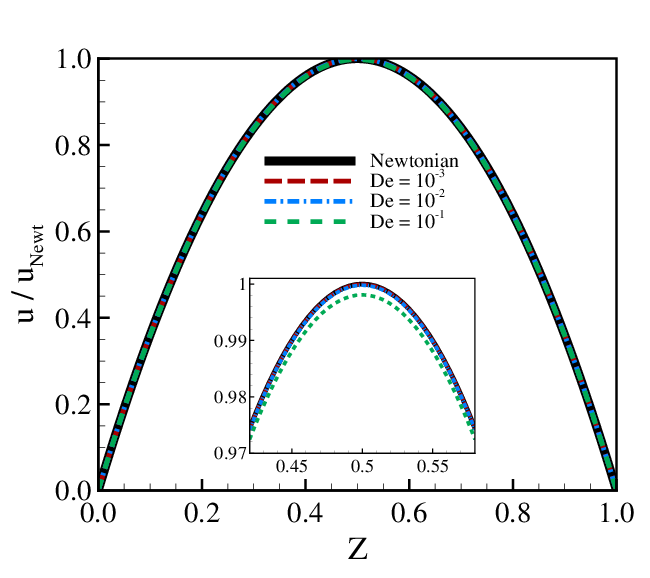}
\caption{Normalised radial velocity profile for a Newtonian fluid and a viscoelastic fluid characterised by the Oldroyd-B model under $De = \{10^{-3}, 10^{-2}, 10^{-1}\}$ and $\beta = 0$.}	
\label{Fig:velocityDe}
\end{figure}

Solving (\ref{Eq:Mom_rAdim_s}) under the boundary conditions in (\ref{Eq:CF}) in proper dimensionless form, the dimensionless radial velocity profile as a function of radial and axial directions, $u(R,Z)$, is:
 
\begin{equation}
u(R,Z) = \frac{1}{\alpha} Ln \left( \frac{e^{\omega(Z_1(R)+Z_2(R)-Z)}}{e^{\omega Z_1(R)} + e^{\omega Z_2(R)}} + \frac{e^{\omega(Z_1(R)+ Z)} + e^{\omega(Z_2(R)+ Z)}}{\left( e^{\omega Z_1(R)} + e^{\omega Z_2(R)} \right)^2 } \right),
\label{Eq:distu}
\end{equation} 
\
where $\alpha = \frac{2(1-\beta)De}{R}$ and $\omega = \sqrt{-\alpha \frac{\partial P}{\partial R}}$, and whose solution step-by-step is provided in Appendix A. Here, the pressure-gradient is assumed to be negative and is yet to be determined. The velocity profile in expression (\ref{Eq:distu}) draws a parabolic functionality at very low \textit{De}-numbers; however, due to the first normal-stress difference ($N_1$) and geometric changes through $\left( \frac{N_1}{R} \right)$ in (\ref{Eq:Mom_rAdim_s}), the velocity profile is reduced in the centre of the gap between the spheres.
Figure \ref{Fig:velocityDe} shows the influence of viscoelasticity in the radial velocity profile under steady-state. We introduce a normalised velocity, defined as $\frac{u}{u_{Newt}}$ that represents the ratio of the viscoelastic velocity profile with respect to Newtonian velocity profile under the same pressure-gradient. The Newtonian velocity profile is recovered from solution (\ref{Eq:distu}) when taking $De = 0$ and $\beta = 1$, such that:

\begin{equation}
u_{Newt}(R,Z) = \frac{1}{2}\frac{\partial P}{\partial R}(Z-Z_1(R))(Z-Z_2(R)).
\label{Eq:distuN}
\end{equation}
\
A viscoelastic solution characterised by $De = 10^{-6}$ behaves close to the Newtonian case. By increasing viscoelasticity to $De = \{10^{-3}, 10^{-2} \}$, solutions display negligible variation with respect to the Newtonian case along the axial direction. However, for the $De = 10^{-1}$ case, a slight difference is apparent around the maximum velocity (see inset in Figure \ref{Fig:velocityDe}), which lies below the Newtonian profile. This result suggests that viscoelasticity slows down the velocity at the centreline for the viscoelastic case, this being relative to the Newtonian profile under the same pressure-gradient.

\subsection{Flow rate and mass conservation} \label{Sec:flowrate}
The flow rate $q$ arises when two approaching spheres squeeze the fluid trapped between them, and is provoked by the relative motion of the particles. The dimensionless flow-rate measure $Q$ is defined as follows:

\begin{equation}
Q = \frac{q}{\pi a^2 V} = 2 \int_{Z_1(R)}^{Z_2(R)} u(R,Z) R \; dZ.
\label{Eq:flow1}
\end{equation}
\	
The expression (\ref{Eq:flow1}) indicates that the flow-rate is a function of the radial direction; therefore, the force driving the velocity profile, i.e., the pressure-gradient in this case, must have a determined value in order to satisfy the mass conservation principle. Integrating the mass conservation equation  \citep{HoriBOOK2006,RuangStonePRF2024} in (\ref{Eq:Cont_Adim}) under the boundary condition in (\ref{Eq:CF}), in
proper dimensionless form, i.e. $w(Z_1) = 0$ and $w(Z_2) = -1$, yields:

\begin{equation}
\int_{Z_1(R)}^{Z_2(R)} u R \; dZ - \frac{R^2}{2} = 0.
\label{Eq:Cont_int}	
\end{equation}
\\
Multiplying (\ref{Eq:Cont_int}) by $2$, it may be re-cast as:
\begin{equation}
2\int_{Z_1(R)}^{Z_2(R)} u R \; dZ - R^2 = Q - R^2 = 0.
\label{Eq:Cont_mas}
\end{equation}
\
Then, (\ref{Eq:Cont_mas}) establishes the mass-conservation condition that will be employed for subsequent calculations (see Appendix B for further details). The steps to determine the velocity profile, the pressure-gradient and the solution of (\ref{Eq:Cont_mas}) at each radial position, from the centreline ($R_i=0$) to the edge of the spheres ($R_i=a$), assuming spheres of the same radius $\left( \kappa_a = \frac{1}{2} \right)$, are given as follows:

\begin{itemize}
\item Firstly, for a given radial position ($R_i$), two estimates of $\frac{\partial P}{\partial R}_i$ are proposed from the Newtonian solution, i.e., $\frac{\partial P}{\partial R}_{New} = \frac{-6R}{H^3(R)}$, which is equivalent to the expression given in \cite{VazquezElleroPF2016} in dimensionless form. 
\
\item Secondly, $Q$ is calculated for both estimates. Here, the integral on the LHS of (\ref{Eq:flow1}) is determined by applying the Simpson’s 3/8 rule.
\
\item Finally, the secant method is used to determine a new estimate for $\frac{\partial P}{\partial R}_i$. This method is employed until $|Q\left( \frac{\partial P}{\partial R}_i \right) - R_i^2 | < 10^{-6}$.
\end{itemize}

\begin{figure}
\centering
\begin{minipage}{0.49\textwidth}
\centering
\includegraphics[width=\linewidth]{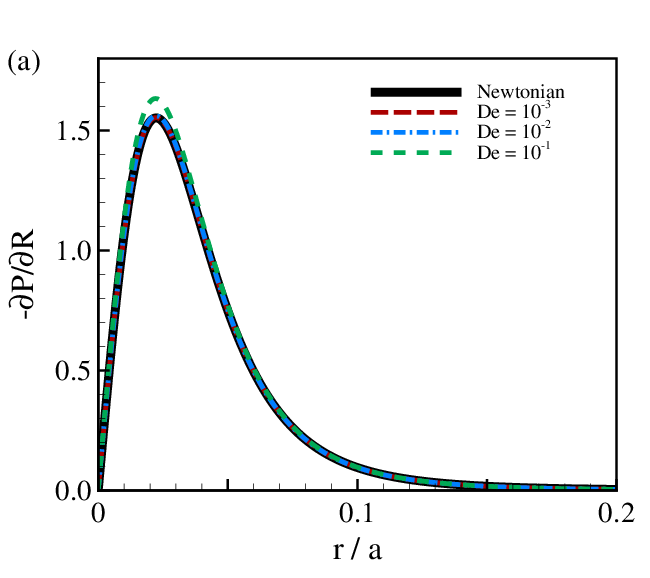}
\end{minipage}
\hfill
\begin{minipage}{0.49\textwidth}
\centering
\includegraphics[width=\linewidth]{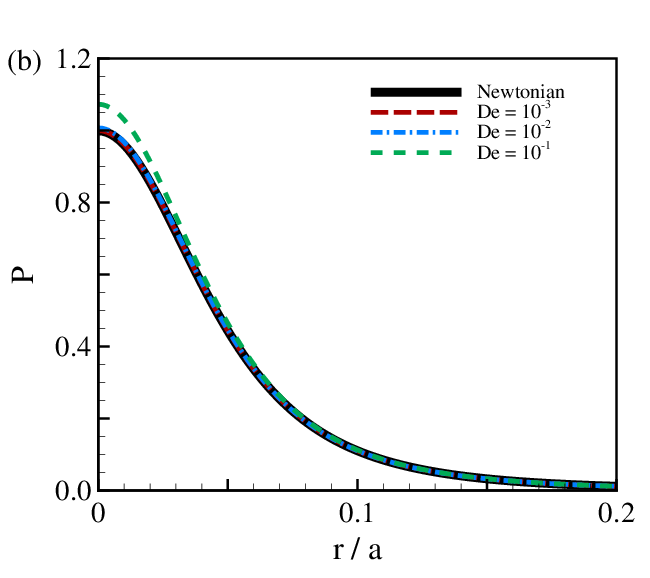}
\end{minipage}
\caption{Numerical solutions for: (a) the pressure-gradient (Simpson’s 3/8 rule and secant method) and (b) pressure (Euler method) under the viscoelastic parametrisation: $\beta = 0$, $a = 1$, $h_0 = 2.5 \times 10^{-3}$, $De = \{10^{-3}, 10^{-2}, 10^{-1} \}$, and under the same sphere approach velocity.}	
\label{Fig:pressureDe}
\end{figure}

The pressure-gradient profile in the radial direction is illustrated in Figure \ref{Fig:pressureDe}(a). The pressure gradient shows a maximum near the centreline, beyond which it drops steeply along the radial direction. The pressure-gradient increases slightly for $De = \{10^{-3}, 10^{-2}\}$; however, for $De = 10^{-1}$, it can be up to $\approx 4\%$ higher than in the Newtonian case at the peak.

Then, the pressure can be approximated by integrating the pressure-gradient from Figure \ref{Fig:pressureDe}(a) within integration limits ranging from $R=0$ to $R = a$, using Euler's method, to obtain:

\begin{equation}
P = - \int_{R}^{\infty} \frac{\partial P}{\partial R} \; d R \approx - \int_{R}^{a} \frac{\partial P}{\partial R} \; d R.
\label{Eq:Pressure}
\end{equation}
\
Pressure attains its maximum in the centreline, located at $R=0$, and declines rapidly within a small radius range, as illustrated in Figure \ref{Fig:pressureDe}(b). Solutions under small \textit{De}-numbers $De=\{10^{-3},10^{-2}\}$ appear similar to the Newtonian solution. Only under $De = 10^{-1}$, there is an evident deviation from the Newtonian behaviour, with pressures some $\approx 6\%$ higher than in the Newtonian case at the peak. These findings may be explained from an analysis of (\ref{Eq:Mom_rAdim_s}) and (\ref{Eq:Cont_mas}), where one finds an additional force originating from viscoelasticity and the first normal-stress difference. This is correlated with a slight reduction in the velocity profile (Figure \ref{Fig:velocityDe}), which consequently leads to a decrease in flow rate as described in (\ref{Eq:flow1}). Therefore, to comply with mass conservation (\ref{Eq:Cont_mas}), there is a corresponding rise in both pressure gradient and pressure.

\subsection{Stress and normal forces}

\begin{figure}
\centering
\includegraphics[width=0.6\linewidth]{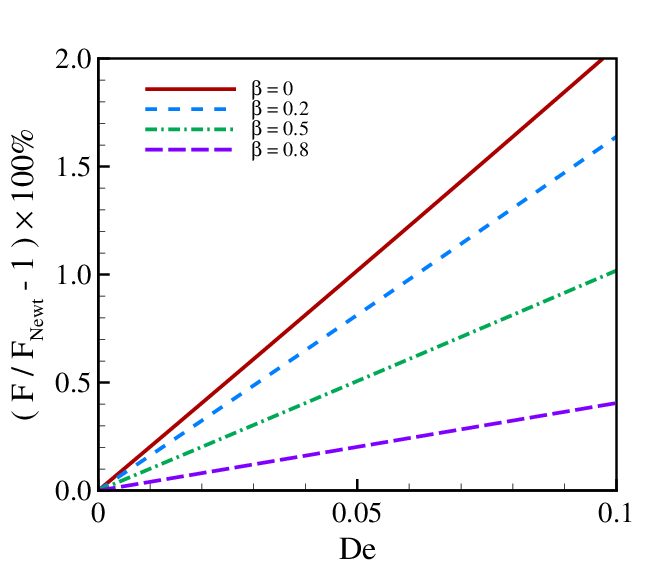}
\caption{Percent increase on the viscoelastic force with respect to a Newtonian response, against \textit{De}-number under $\beta=\{0, 0.2, 0.5, 0.8\}$.}	
\label{Fig:ForceDeNew}
\end{figure}

The normal force acting on the particles' surface ($f)$ is therefore obtained by integration of the total stresses along their surface. Again, Euler's method is employed to calculate the force from the numerical approximation of pressure in the range of $R=0$ to $R = a$. The dimensionless net viscoelastic lubricating force ($F$) normal to the approach-velocity direction on the particle surface $S_2$ is given by:

\begin{equation}
F = \frac{f}{6\pi \eta_0 V a}= \frac{1}{3 \epsilon} \int_0^{\infty} P\mid_{Z = Z_2(R)} R \; d R \approx  \frac{1}{3 \epsilon} \int_0^{a} P\mid_{Z = Z_2(R)} R \; d R.
\label{Eq:force_exp}
\end{equation}
\
Figure \ref{Fig:ForceDeNew} compares the percent increase of the viscoelastic lubrication force against that of Newtonian lubrication force; such Newtonian measure is defined in dimensionless form as: $F_{Newt} = \frac{1}{4 \epsilon}$. As illustrated in Figure \ref{Fig:ForceDeNew}, the normal lubrication force exerted on the $S_2$-particle surface  increases weakly due to the increase in pressure gradient caused by viscoelasticity. Notably, the increasing presence of the Newtonian solvent in the dissolution, measured under $\beta$-increase, reduces the force intensity. For $De = 1 \times 10^{-1}$ and $\beta = 0$, the viscoelastic lubrication force shows the largest rise relative to the Newtonian lubrication force, reaching an increase of up to $\approx 2 \%$. One should bear in mind that in a steady state, the lubrication force exerted on the particles acts immediately. As discussed, the incorporation of viscoelasticity into the momentum balance equation via $N_1$, results in minor effects in the velocity, pressure-gradient, and pressure distributions. Thus, the effect of viscoelasticity on the lubrication force is small when the fluid response is considered instantaneous and when the effects of geometric changes derived from the particle approach are not considered.

\section{Transient case} \label{Sec:Transient}

\subsection{Transient analysis}

In the previous section, the steady-state case demonstrates the instantaneous response of viscoelastic fluids when subjected to a constant squeezing velocity. In this section, we introduce a time-dependent scenario, with the objective of examining the interplay between the particle dynamics and the time response of viscoelastic fluids characterised with the Oldroyd-B model. Therefore, we analyse the transient case by retaining the time-dependent terms in the stress components (\ref{Eq:Tprr})-(\ref{Eq:Tpzz}), which renders the following partial differential-equation system:

\begin{equation}
S_{rr} + De \left( \frac{\partial S_{rr}}{\partial T} - 2 \frac{S_{zr}}{\epsilon^{1/2}}  \frac{\partial u}{\partial Z} \right) = 0, 
\label{Eq:TprrAdim_t}
\end{equation} \
\begin{equation}
S_{rz} + De \left( \frac{\partial S_{rz}}{\partial T} - \frac{S_{zz}}{\epsilon^{1/2}} \frac{\partial u}{\partial Z} \right) = (1 - \beta) \frac{\partial u}{\partial Z},
\label{Eq:TprzAdim_t}
\end{equation} \
\begin{equation}
S_{\theta\theta} + De \frac{\partial S_{\theta\theta}}{\partial T} = 0,
\label{Eq:TpttAdim_t}
\end{equation} \
\begin{equation}
S_{zz} + De \frac{\partial S_{zz}}{\partial T} = 0.
\label{Eq:TpzzAdim_t}
\end{equation} 
\
For the corresponding initial conditions, we consider a completely relaxed stress, as follows:
\
\begin{equation}
\boldsymbol{S} = \boldsymbol{0}, \; \text{for } T = 0.
\label{Eq:CI}
\end{equation}

As particle $S_2$ moves with constant velocity $V$ towards  particle $S_1$, its contour location changes with time, such that: $z_2(r,t) = h_0 + \frac{r^2}{2a} - Vt$. Therefore, the dimensionless interparticle distance reduces with time as follows: 

\begin{equation}
H(R,T) = \frac{z_2(r,t) - z_1(r)}{h_0} = \frac{h(r,t)}{h_0} = 1 + \frac{R^2}{2 \kappa_a} - T.
\label{Eq:HRT}
\end{equation}
\
From here, the dimensionless time required for the two spheres to come into contact at $H_0 = H(0,T) = 0$, is $T = 1$.

\subsection{Transient solutions} \label{sSec:TrasientSol}

Solving for (\ref{Eq:TpttAdim_t})-(\ref{Eq:TpzzAdim_t}) under the initial condition in (\ref{Eq:CI}), one obtains:

\begin{equation}
S_{\theta \theta} = 0,
\label{Eq:Spoo_sol}
\end{equation} 
\begin{equation}
S_{zz} = 0.
\label{Eq:Spzz_sol}
\end{equation}
\
Substituting (\ref{Eq:Spzz_sol}) in (\ref{Eq:TprzAdim_t}) and solving for the dimensionless shear stress under the initial condition in (\ref{Eq:CI}), yields:

\begin{equation}
S_{rz} = ( 1 - \beta)\frac{\partial u}{\partial Z} \left( 1 - e^{-T/De} \right).
\label{Eq:Sprz_sol}
\end{equation}
\
Then, substituting (\ref{Eq:Sprz_sol}) in (\ref{Eq:TprrAdim_t}) and solving for the normal stress in the radial direction under initial condition in (\ref{Eq:CI}), renders:
\begin{equation}
S_{rr} = \frac{2(1 - \beta)}{\epsilon^{1/2}} \left( \frac{\partial u}{\partial Z} \right)^2 \left(  De - \left( De + T \right) e^{-T/De} \right).
\label{Eq:Sprr_sol}
\end{equation} 

Defining: $g(T) = \left( 1 - e^{-T/De} \right)$ and $f(T) = \left( De - \left( De + T \right) e^{-T/De} \right)$, the velocity profile is calculated substituting the shear stress in (\ref{Eq:Sprz_sol}) and the normal stress in (\ref{Eq:Sprr_sol}) in the motion equation, (\ref{Eq:Mom_rAdim_N}), as follows:
\
\begin{equation}
- \frac{\partial P}{\partial R} + \beta \frac{\partial^2 u}{\partial Z^2} + \frac{\partial}{\partial Z} \left( ( 1 - \beta) g(T)\frac{\partial u}{\partial Z} \right) + \frac{2 (1 - \beta) f(T)}{R} \left( \frac{\partial u}{\partial Z} \right)^2 = 0.
\label{Eq:pODE}
\end{equation}
\\
Differentiation with respect to $Z$ of (\ref{Eq:pODE}), grouping of similar terms, and considering the following identities:
\
\begin{equation}
\alpha(T) = \frac{2 (1 - \beta) f(T)}{R \left( \beta + ( 1 - \beta) g(T) \right) }, \quad \frac{\partial P'}{\partial R}(T) = \frac{1}{\beta + ( 1 - \beta) g(T)} \frac{\partial P}{\partial R},
\end{equation}
\
the momentum balance equation in $r-$direction results in:
\begin{equation}
- \frac{\partial P'}{\partial R}(T) + \frac{\partial^2 u}{\partial Z^2} + \alpha(T) \left( \frac{\partial u}{\partial Z} \right)^2 = 0.
\label{Eq:ODEu}
\end{equation}
\
The momentum equation in (\ref{Eq:ODEu}) relies on temporal functions $\left( \alpha(T), \frac{\partial P'}{\partial R}(T) \right)$ that come from the viscoelastic stress. These terms emerge from the time-dependent development of the velocity field, as $S_2$ contour relocates over time. 
Integration of (\ref{Eq:ODEu}) under the boundary conditions in (\ref{Eq:CF}), yields the following velocity profile:

\begin{equation}
u(R,Z,T) = \\
\frac{1}{\alpha} Ln \left( \frac{e^{\omega(T)(Z_1(R)+Z_2(R,T)-Z)}}{e^{\omega(T) Z_1(R)} + e^{\omega(T) Z_2(R,T)}} + \frac{e^{\omega(T)(Z_1(R)+ Z)} + e^{\omega(T)(Z_2(R,T)+ Z)}}{\left( e^{\omega(T) Z_1(R)} + e^{\omega(T) Z_2(R,T)} \right)^2 } \right),
\label{Eq:uprofile_t}
\end{equation}
\
where, $\omega(T) = \sqrt{-\alpha(T) \frac{\partial P'}{\partial R}(T)}$. 

\begin{figure}
\centering
\includegraphics[width=0.6\textwidth]{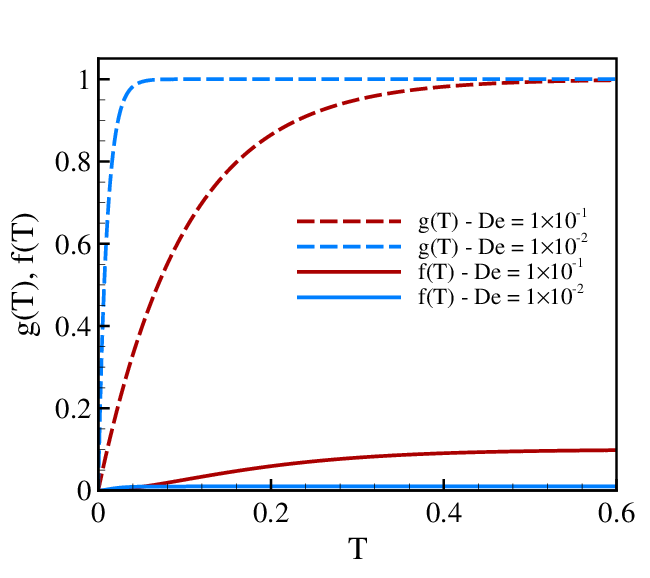}
\caption{Temporal functions $g(T)$ and $f(T)$ originating from viscoelastic stress under $De = \{1 \times 10^{-2}, 1 \times 10^{-1} \}$ and $\beta = 0.5$.}
\label{Fig:temp_func}
\end{figure}

Here, the temporal functions $\left( \alpha(T), \frac{\partial P'}{\partial R}(T) \right)$ have the following limiting values:

\begin{equation}
\alpha(T = 0) = 0, \quad \quad \alpha(T \rightarrow 1) = \frac{2(1 - \beta)De}{R},
\label{Eq:alimits}
\end{equation}
\
\begin{equation}
\frac{\partial P'}{\partial R}(T = 0) = \frac{1}{\beta} \frac{\partial P}{\partial R}, \quad \frac{\partial P'}{\partial R}(T \rightarrow 1) = \frac{\partial P}{\partial R}.
\label{Eq:glimits}
\end{equation}
\
From the these limits in (\ref{Eq:alimits}) and (\ref{Eq:glimits}), one may conclude that:

\begin{itemize}
\item At $T = 0$, there is an instantaneous viscous response described by:
\begin{equation}
- \frac{\partial P}{\partial R} + \beta \frac{\partial^2 u}{\partial Z^2} = 0,
\label{Eq:ODEu_0}
\end{equation}
\
which comes from (\ref{Eq:ODEu}) when viscoelastic effects are null, and represents the force balance for a Newtonian fluid of viscosity $\eta_0 = \eta_s$. Therefore, the viscoelastic behaviour starts from a Newtonian fluid response.

\item In the range $0 < T < 1$, the viscoelastic fluid response may evolve from a Newtonian fluid of viscosity $\eta_0 = \eta_s$ to a viscoelastic fluid of viscosity $\eta_0 = \eta_s + \eta_p$, whose force balance is given by (\ref{Eq:ODEu}). The dimensionless time expected to reach a steady value of $(g(T), f(T))$ is around $T \sim 0.5$ even for $De = 1 \times 10^{-1}$ (see Figure \ref{Fig:temp_func}). However, since $Z_2(R,T)$ varies with time, the distance between the spheres will diminish and generate geometric changes \citep{RuangStonePRF2024}, indicating that they are moving closer towards touching each other.

\item In the limit of $T \rightarrow 1$, (\ref{Eq:glimits}) simplifies to a version resembling the steady-state scenario in (\ref{Eq:Mom_rAdim_s}), that is, a viscoelastic fluid with a viscosity of $\eta_0 = \eta_s + \eta_p$. Here, the contour of sphere $S_2$ continues to draw nearer to sphere $S_1$, which promotes the fluid to get confined.
\end{itemize}
\
\
Therefore, the response at relatively short times resembles the response of a Newtonian fluid with viscosity $\eta_s$, indicating that the parameter $\beta$ modulates the initial lubrication force. In the case of an upper-convected Maxwell fluid, i.e., for $\beta = 0$, the force will start from zero, and, for $\beta = 1$, the force will align with the expected response of a Newtonian fluid. The computations for pressure-gradient, pressure, and normal lubrication force follow the same numerical techniques outlined in Section \ref{Sec:flowrate},  adding the functionality with time through $Z_2(R,T)$, $g(T)$ and $f(T)$. Considering the time in the domain $(0 \leq T < 1)$ in the transient case, the viscoelastic lubrication force is expected to increase over time or increase as distance is reduced, according with (\ref{Eq:HRT}). In order to detect a significant alteration in distance, time needs to exceed $0.9$, such that $T > 0.9$.

\begin{figure}
\centering
\includegraphics[width=0.6\textwidth]{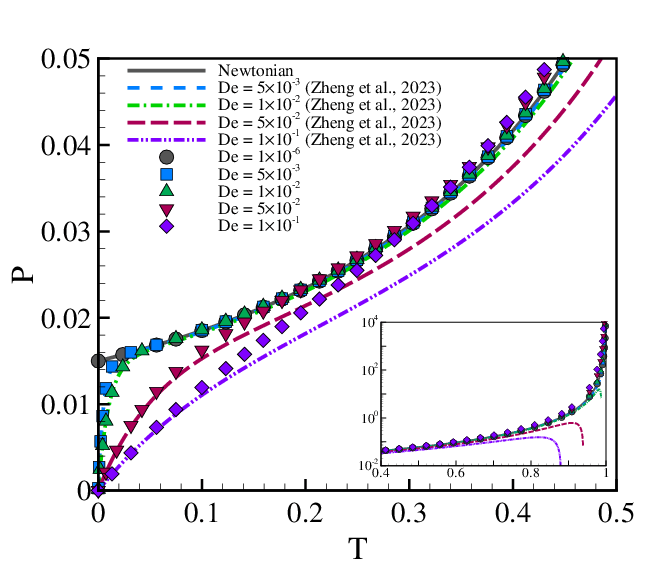}
\caption{Dimensionless pressure against time under $De = \{1 \times 10^{-6}, 5 \times 10^{-3}, 1 \times 10^{-2}, 5 \times 10^{-2}, 1 \times 10^{-1} \}$ for $\epsilon = 2.5 \times 10^{-3}$. Symbols identify our numerical solutions whilst lines come from Eq.(52) in \cite{ZhengetalPF2023}.}
\label{Fig:press_trans}
\end{figure}

We compare our solutions with expressions (52) and (54) in \cite{ZhengetalPF2023}, corresponding to the pressure and the lubrication force, respectively. The main distinction between our formulation and that of \cite{ZhengetalPF2023}, lies in their prescription of a Newtonian velocity profile, a stress-based boundary condition to compute the pressure-gradient, and a fluid characterised with the Maxwell model is considered. Figure \ref{Fig:press_trans} displays the time evolution of the \textit{pressure} for different values of \textit{De}-numbers comparing our solutions to those obtained by \cite{ZhengetalPF2023}. For early times, specially under $T \lesssim 0.3$, our predictions yield pressures lower than those of the Newtonian reference solution. However, for $T \gtrsim 0.3$, the viscoelastic predictions converge to, and marginally exceed, the Newtonian result. In that sense, the development of the viscoelastic time-dependent functions in (\ref{Eq:ODEu}) promotes a rise both pressure and lubrication force. Noteworthy is the divergence of our solutions with respect to those of \cite{ZhengetalPF2023}, which become apparent as $T \rightarrow 1$.  \cite{ZhengetalPF2023} model predicts a decline in pressure, approaching negative values (see the inset in Figure \ref{Fig:press_trans}), in contrast to the monotonic increase characteristic of the Newtonian solution, which is predicted with our model.

\begin{figure}
\centering
\includegraphics[width=0.6\linewidth]{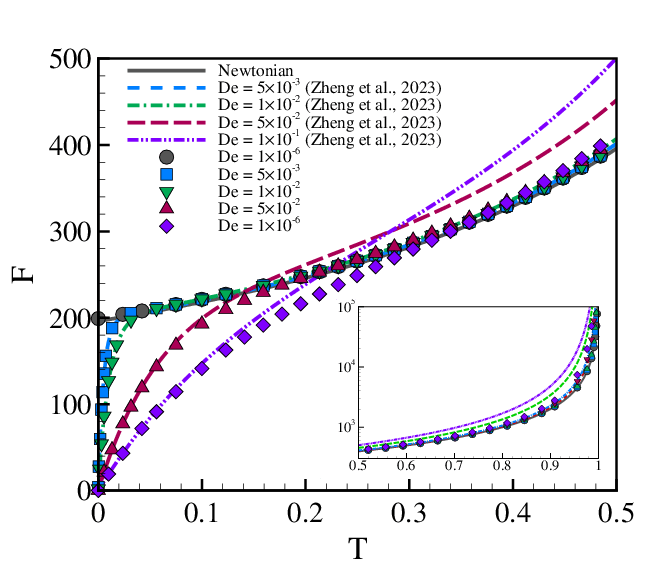}
\caption{Dimensionless lubrication force against time under $De = \{1 \times 10^{-6}, 5 \times 10^{-3}, 1 \times 10^{-2}, 5 \times 10^{-2}, 1 \times 10^{-1} \}$ for $\epsilon = 2.5 \times 10^{-3}$. Symbols identify our numerical solutions whilst lines come from Eq.(52) in \cite{ZhengetalPF2023}.}	
\label{Fig:forceDe_Zheng}
\end{figure}

For the \textit{normal lubrication force} in Figure \ref{Fig:forceDe_Zheng}, our solutions are in close agreement with those by \cite{ZhengetalPF2023} at short times. In addition, both models follow the Newtonian upper limit for all \textit{De}-numbers. Nonetheless, for $T \gtrsim 0.2$, \cite{ZhengetalPF2023} model predicts a lubrication force that exceeds our solutions for equivalent \textit{De}-levels.

\section{Model predictions} \label{Sec:ModelPre}

\begin{figure}
\centering
\includegraphics[width=0.6\linewidth]{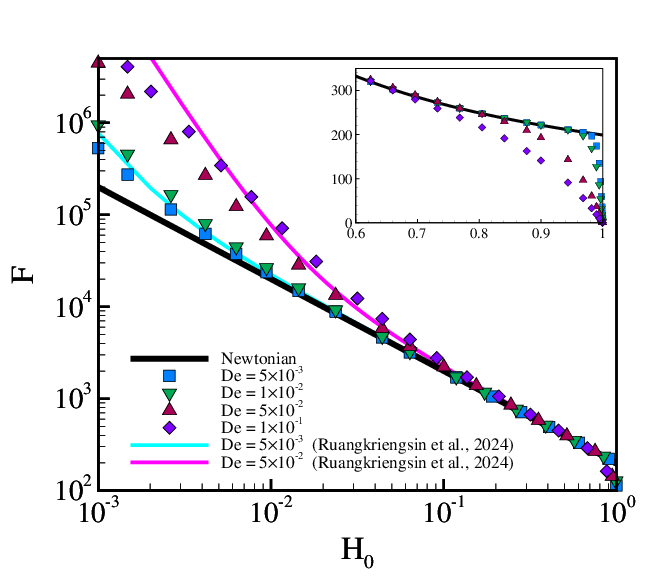}
\caption{Dimensionless viscoelastic lubrication force against interparticle distance under $De = \{5 \times 10^{-3}, 1 \times 10^{-2}, 5 \times 10^{-2}, 1 \times 10^{-1} \}$, $\beta = 0$, and $\epsilon = 2.5 \times 10^{-3}$ and two cases extracted from \cite{RuangStonePRF2024} model.}
\label{Fig:forceDe_H0}
\end{figure}

Whilst comparison of our predictions against those by \cite{ZhengetalPF2023} provides insights into the behaviour of pressure and normal lubrication forces over time, it is typical to represent the lubrication force relative to the distance between the spherical contours. This can be achieved by evaluating (\ref{Eq:HRT}) at various time intervals. In Figure \ref{Fig:forceDe_H0}, results are illustrated for a reference Newtonian fluid and for different viscoelastic fluids described by the Oldroyd-B model. The reference Newtonian fluid displays a linear declining trend with $H_0$-rise,  whilst for the Oldroyd-B model, a linear trend is observed only for interpartle gaps in the range $H_0 \approx (0.1 - 0.6)$. Notably,  in the interval $H_0 = (0.6, 1]$, the viscoelastic lubrication force declines sharply to zero (see inset of Figure \ref{Fig:forceDe_H0}).  In contrast, for shorter distances, i.e., $H_0 \lesssim 0.1$, the viscoelastic lubrication force deviates from the Newtonian reference-line,  increasing with $H_0$-decrease. This enhancement of the viscoelastic lubrication force with reducing interparticle gaps is detailed in Table \ref{Tb:Flub}. Notably, even at the smallest \textit{De}-number examined, i.e., $De=5 \times 10^{-3}$, this increase can be as high as $70\%$ over the Newtonian case, whilst for the largest \textit{De}-number recorded, i.e., $De=1 \times 10^{-1}$,  this viscoelastic lubrication-force enhancement can amount over three orders-of-magnitude larger than the corresponding Newtonian instance, with a $2113\%$ excess. This rise in lubrication force for viscoelastic fluids under transient-confinement dynamics, evidenced here using the Oldroyd-B model, could be crucial for modelling dense systems with particles suspended in a viscoelastic medium with constant shear-viscosity Boger-fluid features. It is important to bear in mind that we have considered constant \textit{De} numbers in our calculations. However, as spheres approach, the interparticle gap reduces,  and, consequently,  \textit{De} increases (see (\ref{Eq:adimvar})),  possibly leading to further strengthening this viscoelastic enhancement on the normal lubrication force. In addition, we compare our solutions against the predictions for a sphere moving towards a rigid plane by \cite{RuangStonePRF2024} for $De = \{ 5 \times 10^{-3}, 5 \times 10^{-2} \}$.  Here, our predictions and those by \cite{RuangStonePRF2024} concur in trends, only deviating at relatively reduced interparticle gaps and increased \textit{De} numbers.

\begin{table}
    \begin{center}
    \def~{\hphantom{0}} 
    \begin{tabular}{ccccc}
    \hline
    \multirow{2}{*}{$H_0$} & \multicolumn{4}{c}{$De$} \\ \cline{2-5} 
    & {$5 \times 10^{-3}$} & {$1 \times 10^{-2}$} & {$5 \times 10^{-2}$} & {$1 \times 10^{-1}$} \\
    $1 \times 10^{-1}$ & $1\%$ & $2\%$ & $11\%$ & $24\%$ \\
    $1 \times 10^{-2}$ & $9\%$ & $20\%$ & $138\%$ & $314\%$ \\
    $2 \times 10^{-3}$ & $70\%$ & $163\%$ & $1018\%$ & $2113\%$ \\
    \hline
    \end{tabular}
    \caption{Percent increase of the viscoelastic lubrication force over the Newtonian reference for a set of \textit{De}-numbers and dimensionless interparticle distances.} \label{Tb:Flub}
    \end{center}
\end{table}

\subsection{Effect of the first normal stress-difference} \label{Sec:NormalS}

\begin{figure}
\centering
\begin{minipage}{0.49\textwidth}
\centering
\includegraphics[width=\linewidth]{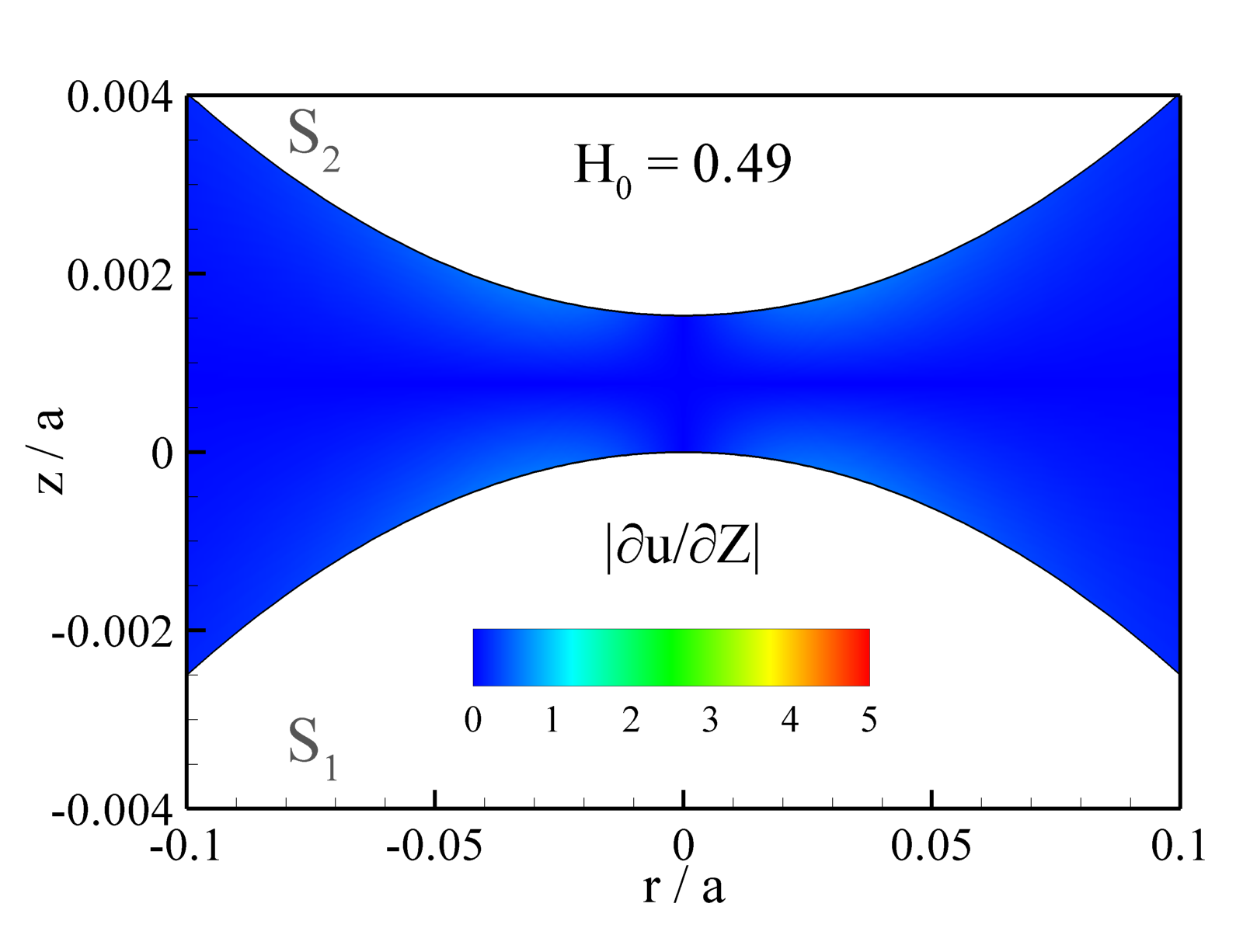}
\end{minipage}
\hfill
\begin{minipage}{0.49\textwidth}
\centering
\includegraphics[width=\linewidth]{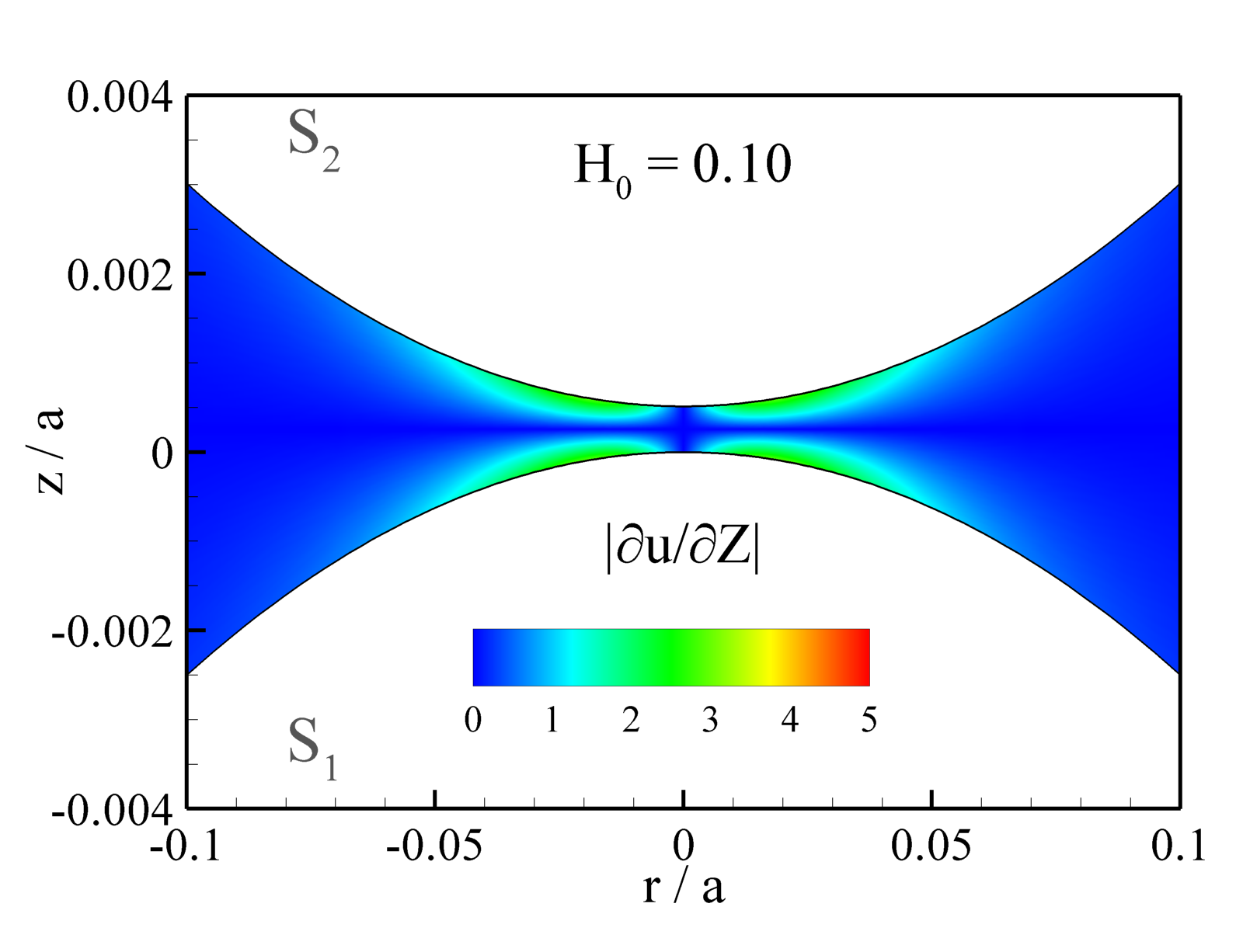}
\end{minipage}
\begin{minipage}{0.49\textwidth}
\centering
\includegraphics[width=\linewidth]{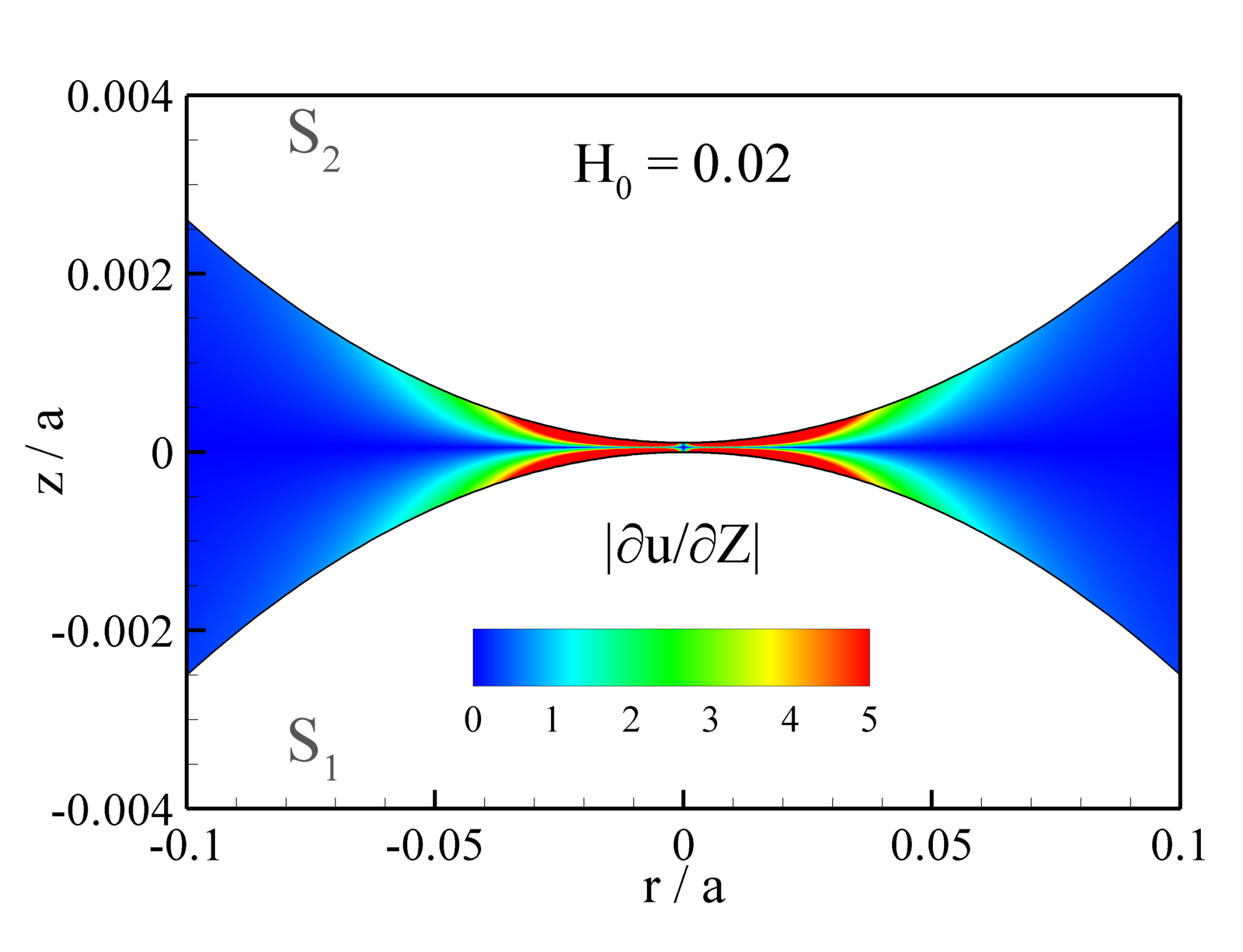}
\end{minipage}
\hfill
\begin{minipage}{0.49\textwidth}
\centering
\includegraphics[width=\linewidth]{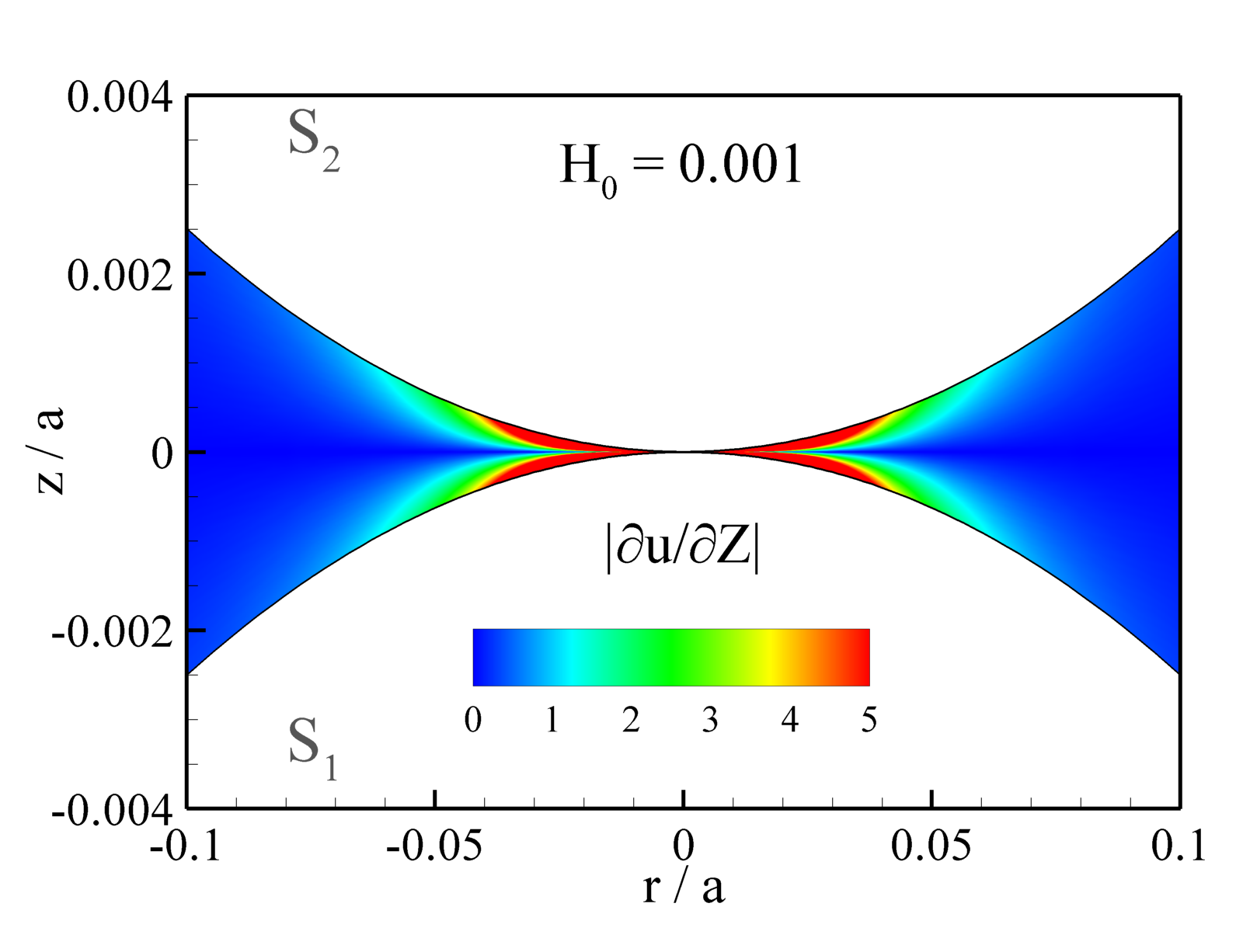}
\end{minipage}
\caption{Dimensionless velocity gradient in the interparticle separation between the spheres for a viscoelastic fluid under $De = 5 \times 10^{-2}$, $\beta = 0$, $a = 1$, and $h_0 = 2.5 \times 10^{-3}$.}	
\label{Fig:dvrdz}
\end{figure}

To further describe the viscoelastic lubrication-force enhancement promoted by fluid confinement exposed in previous paragraphs, we examine the topology of the velocity gradient in the interstitial fluid separating both spheres. As outlined in Section \ref{Sec:Steady}, the viscoelastic component of stress in the momentum balance is represented by the first normal-stress difference,  particularly contained in the term: $\alpha(T) \left( \frac{\partial u}{\partial Z} \right)^2$ (see (\ref{Eq:ODEu}) and the definition of the radial normal stress $S_{rr}$), which denotes an extra resistance to the shearing deformation of the fluid during the constant-velocity sphere-approaching modality. This lubrication-force enhancement may emerge through this extra term due to viscoelasticity, which scales with the square of the shear-rate. In addition, the major contribution to this viscoelastic resistance occurs near the centre of the particle gap. Figure \ref{Fig:dvrdz} illustrates the velocity-gradient magnitude in the separation between the spheres over time; here,  the blue colour-shading represents a relatively small shear-rate magnitude, whilst the red colour-shading depicts relatively large shear-rate values. Under the relatively large interparticle gaps of $H_0 = 0.49$, the velocity gradient across the space between particles remains minimal. At $H_0 = 0.10$, there is a slight increase in the velocity-gradient magnitude on the surface of the spheres near the centre, with magnitude around $\sim 1$ units, which leads to a minor rise in the lubrication force. As the interparticle gap decreases further in one order-of-magnitude up to $H_0 = 0.02$, the velocity-gradient magnitude rises moderately to a peak value of $\sim 30$ units. For the significantly-confined instance of $H_0 = 0.001$, the velocity-gradient magnitude increases abruptly up to $\sim 3000$ units.  Given that the viscoelastic lubrication-force enhancement is proportional to the square of the velocity-gradient magnitude, even under small \textit{De}-values, the velocity-gradient magnitude at $H_0 = \{0.02, 0.001 \}$ is large enough to significantly enhance fluid resistance to be squeezed under a constant approaching velocity. This, in turn, and under such settings, increases the pressure gradient needed to comply with mass conservation, and results in a stark enhancement in the normal lubrication force near contact.

\subsection{Boger fluid problem}

To further explore the predictive capabilities of our model, we consider the rheometrical data for the S1 fluid (a dissolution of polyacrylamide in corn syrup with composition: $0.05\%$ Separan AP 30, $2.4\%$ $H_2O$, and $97.55\%$ Corn Syrup 41N) in \cite{PhanetalJNNM1985}. We characterise this fluid with the Oldroyd-B model, which yields the relevant material functions of $\eta_0 = 55.5 \; Pa \cdot s$, $\eta_s = 15.8 \; Pa \cdot s$, and $\lambda = 5 \; s$. In addition and for back reference, two Newtonian fluids are considered under $\eta_0 = \{15.8, 55.5\} \; Pa \cdot s$.
The transient forces are calculated by evaluating (\ref{Eq:force_exp}) on the particle $S_2$ at different times under a constant squeezing velocity. In addition, we investigate the viscoelastic force when the spheres move away from each other. 
The approaching case starts from $h_0 = 2.5 \times 10^{-4}$ to $h_0 = 2.5 \times 10^{-7}$, meanwhile the separating instance starts from $h_0 = 2.5 \times 10^{-7}$ to $h_0 = 2.5 \times 10^{-4}$, both cases kicking off from rest.

\begin{figure}
\centering
\begin{minipage}{0.49\textwidth}
\centering
\includegraphics[width=\linewidth]{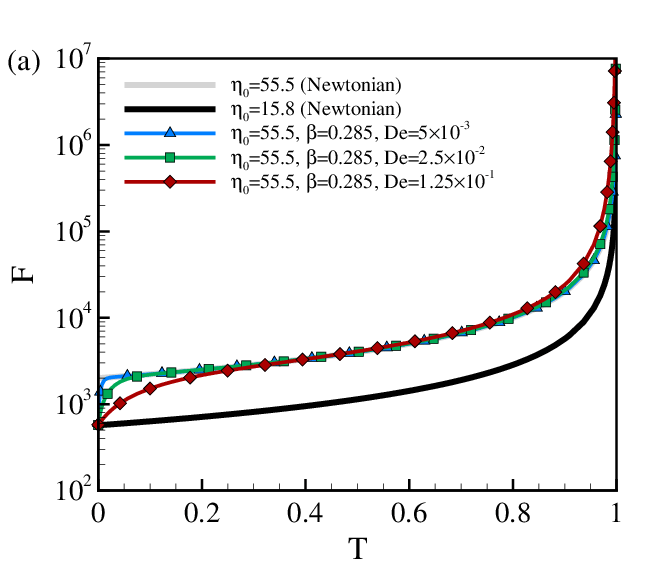}
\end{minipage}
\hfill
\begin{minipage}{0.49\textwidth}
\centering
\includegraphics[width=\linewidth]{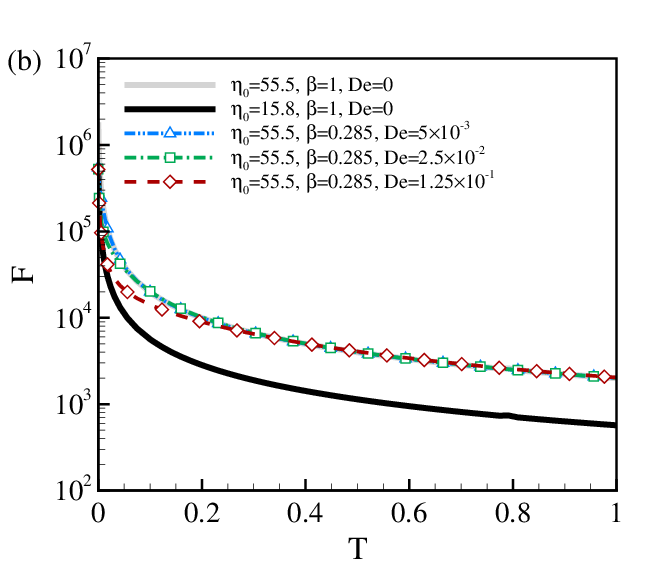}
\end{minipage}
\caption{Force acting on the particle against time for the viscoelastic S1 fluid described in \cite{PhanetalJNNM1985} for different values of \textit{De}-numbers and two Newtonian fluids. For viscoelastic Boger fluids, (a) the sphere-approaching mode is represented by filled symbols, whilst (b) the sphere-separating mode is identified through blank symbols. Newtonian fluids are illustrated by solid lines for both modes.}
\label{Fig:forceDe}
\end{figure}

Figure \ref{Fig:forceDe} shows how the viscoelastic force behaves with varying \textit{De}-numbers over time when utilising the fluid S1 \citep{PhanetalJNNM1985}. As the spheres approach each other (illustrated in Figure \ref{Fig:forceDe}(a)), the lubrication force increases as time progresses, whilst when  spheres separate (in Figure \ref{Fig:forceDe}(b)) the lubrication force decreases with time. For both Newtonian fluids, the lubrication force observed at the end of the approaching case match with the initial lubrication force noted during the separation scenario, and vice versa. Contrastingly, with Boger fluids, the normal lubrication force at the end of the approaching case is larger than the initial lubrication force of the separation scenario. Consistently, the starting viscoelastic lubrication force during the separation starts at a lower level with respect to the end of the approaching case.
During the sphere approach and separation modes, the viscoelastic lubrication force seems constrained by a lower limit corresponding to the Newtonian fluid of viscosity $\eta_0 = \eta_s = 15.8 \; Pa \cdot s$, and an upper limit which corresponds to the Newtonian fluid of viscosity $\eta_0 = \eta_s + \eta_p = 55.5 \; Pa \cdot s$. In the sphere approaching case, for $0.4 \lesssim T < 1$, the viscoelastic force exceeds the Newtonian reference, as reported in \cite{PhanetalJNNM1985}.

\begin{figure}
\centering
\begin{minipage}{0.49\textwidth}
\centering
\includegraphics[width=\linewidth]{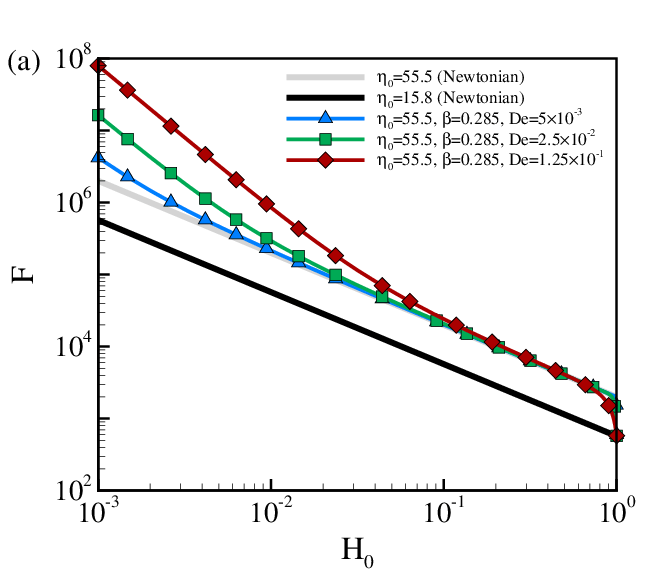}
\end{minipage}
\hfill
\begin{minipage}{0.49\textwidth}
\centering
\includegraphics[width=\linewidth]{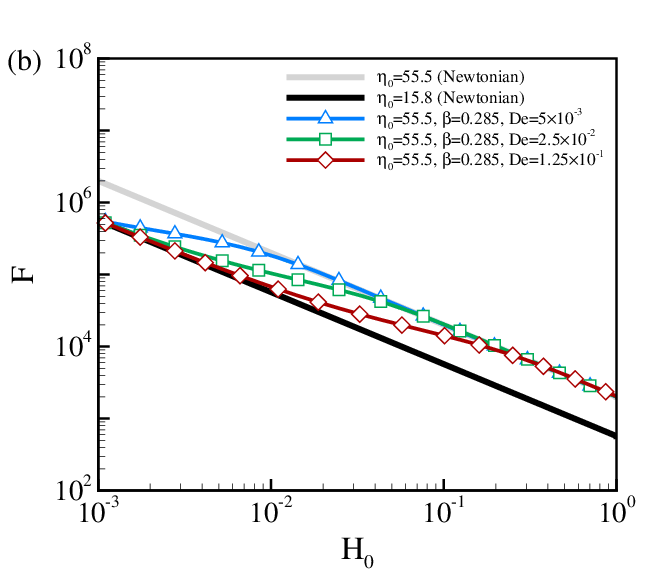}
\end{minipage}
\caption{Dimensionless force on the particle against interparticle distance for the viscoelastic S1 fluid  described in \cite{PhanetalJNNM1985} for different values of \textit{De}-numbers and two Newtonian fluids. For viscoelastic Boger fluids, (a) the sphere-approaching mode is represented by filled symbols, whilst (b) the sphere-separating mode is identified through blank symbols. Newtonian fluids are illustrated by solid lines for both modes.}
\label{Fig:forceH0}
\end{figure}

\begin{figure}
\centering
\includegraphics[width=0.6\linewidth]{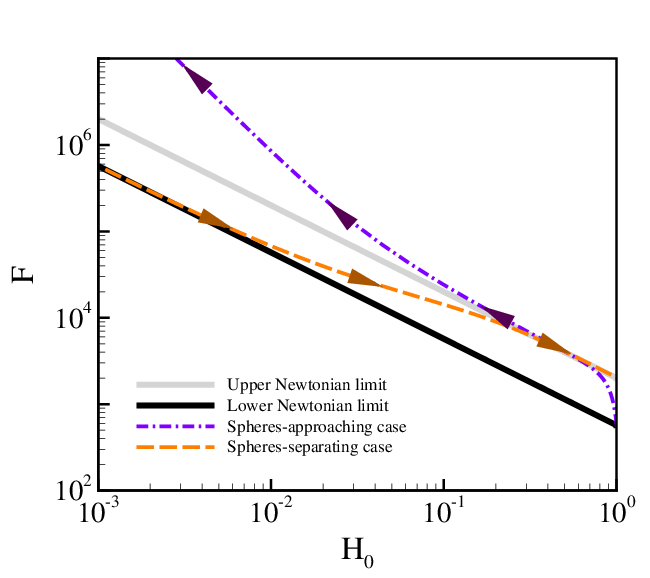}
\caption{Dynamics of the lubrication force for approaching and separating cases for two Newtonian lubrication forces (symmetric response, drawn with continuous lines) and for a Boger fluid characterised with the Oldroyd-B model (asymmetric response, drawn with discontinuous lines).}
\label{Fig:forceH0_C}
\end{figure}

In order to investigate how the evolution relates to distance, we apply the expression $H_0 = 1 - T$ to calculate the dimensionless distance as time advances. Figure \ref{Fig:forceH0} illustrates this situation for dimensionless distances from $H_0 = 1$ to $H_0 = 1 \times 10^{-3}$, which corresponds to times from $T = 0$ to $T = 0.999$. For the sphere-approaching case, the lubrication force increases as the dimensionless distance decreases in time (in Figure \ref{Fig:forceH0}(a); here, the Boger fluid is illustrated with filled symbols, and time advances from right to left). In contrast, under the sphere-separating mode, the lubrication force decreases as distance increases with time (in Figure \ref{Fig:forceH0}(b); here, the Boger fluid is illustrated by hollow symbols, and time advances from left to right). Notably and as a reference behaviour, Newtonian fluids follow the same trajectory under both approaching and separating modes (in Figure \ref{Fig:forceH0}(a) and \ref{Fig:forceH0}(b), Newtonian cases are depicted by solid lines), displaying a symmetric response as a result of the Stokes conditions prevailing.
Contrastingly, for the viscoelastic material in the \textit{approaching scenario}, for distances between $H_0 = 1$ and $H_0 = 1 \times 10^{-1}$, the viscoelastic force remains close to the upper Newtonian threshold. Conversely, when interparticle gaps drop below $H_0 < 1 \times 10^{-1}$ at times $T > 0.9$, there is a significant increase in the magnitude of the viscoelastic force rising up to two orders-of-magnitude for $De = 1.25 \times 10^{-1}$. Under the \textit{sphere-separating instance}, the viscoelastic case maintains its force near the lower Newtonian solution for $H_0 < 10^{-2}$. Beyond such landmark, a shift in the response occurs towards the upper Newtonian solution in the range $1 \times 10^{-2} \leq H_0 \leq 1 \times 10^{-1}$. Hence, one may conclude that viscoelaticity may induce asymmetric responses (see Figure \ref{Fig:forceH0_C}) and may be able to enhance the lubrication force during approaching cases, even assuming small \textit{De}-numbers.

\section{Conclusions} \label{Sec:Conclusions}
In this work, the dynamics of the squeezing flow between two spheres embedded in a viscoelastic fluid,  described by an Oldroyd-B model, is examined. By focusing on the scenario of very narrow gaps between spheres, their shapes can be approximated as paraboloids. This simplification permits the use of lubrication theory to solve the mass and momentum balance equations employing an order-of-magnitude analysis. We consider weak viscoelasticity through small \textit{De}-numbers. Differently to other approaches, we solved for the velocity profile analytically and employed numerical methods to calculate the pressure and normal forces. 

Once the squeezing flow is established, the momentum balance is composed by a viscous response coupled with a viscoelastic response, led by the first normal-stress difference, originating from the polymeric viscoelastic stress. Such momentum balance formulation reproduces a distinct velocity profile from an equivalently-viscous Newtonian profile, fact that evidences the viscoelastic contribution to the flow dynamics.

We examine two flow regimes, i.e., approaching and departing sphere motion under steady and transient conditions. In the \textit{approaching case of steady-state conditions}, representing an immediate system response, viscoelasticity has a marginal effect in the normal lubrication force due to the small \textit{De}-numbers imposed. In this case, the normal lubrication force displays a linear increase with \textit{De}-number and solvent fraction $\beta$.

Conversely, for \textit{transient solutions}, changes in the interparticle gap lead to a time-dependent evolution of normal-stress differences that enhance resistance to flow by up to two orders-of-magnitude compared to the Newtonian response. This resistance is manifested through larger pressure gradients that work as the driving force that ejects the fluid trapped between the spheres, even at the relatively small \textit{De}-numbers imposed. Furthermore, viscoelasticity induces asymmetric responses when spheres approach or separate. These phenomena arising from viscoelasticity may be utilised to modelling the rheological behaviour of dense suspension in viscoelastic Boger fluids, dominated by short-range interactions \citep{KumaretalJNNFM2020,RuizLopezetalJRheol2023tribo,RosalesetalJNNFM2024}.

\subsection*{Acknowledgements}
JEL-A acknowledges the support from Secretar{\'i}a de Ciencia, Humanidades, Tecnolog{\'i}a e Innovaci{\'o}n (SECIHTI, Mexico - grant number CF-2023-I-318), and from Universidad Nacional Autónoma de México (UNAM - grant numbers PAPIIT IN106424 and PAIP 5000-9172 Facultad de Qu{\'i}mica). AR-R acknowledges the support from SECIHTI for the scholarship (CVU 778830) to fund his post-graduate studies.
We acknowledges the funding by the Basque Government through the BERC 2022-2025 program and by the Ministry of Science, Innovation and Universities: BCAM Severo Ochoa Excellence Accreditation CEX2021-001142-S/MICIN/AEI/10.13039/501100011033. 
ME and AV-Q acknowledge the financial support by the Spanish State Research Agency through the project PID2020-117080RB-C55 (‘Microscopic foundations of soft-matter experiments: computational nano-hydrodynamics’) funded by (AEI/FEDER, UE) with acronym COMPU-NANO-HYDRO, and the project PID2020-117080RB-C54 (‘‘Coarse-Graining theory and experimental techniques for multiscale biological systems’’) funded both by AEI – MICIN.

\bibliographystyle{unsrtnat}
\bibliography{REFS_VISCO,REFS}

\appendix
    
\section*{Appendix A. Analytic solution for velocity profile} \label{appA}

Here, the velocity profile derivation is provided. The solution is calculated solving (\ref{Eq:ODEu}) using a reduction-order method,  where $-\frac{d P'}{d R}(T) = \psi$, $\alpha(T) = \alpha$, $\phi = \frac{du}{dZ}$, and $\frac{d\phi}{dZ} = \frac{d^2u}{dZ^2}$. Considering such definitions, (\ref{Eq:ODEu}) becomes:
\begin{equation}
\psi + \frac{d \phi}{d Z} + \alpha \phi ^2 = 0.
\label{Eq:EDO2nL_1}
\end{equation}

Integration and application of the separation-of-variables method in (\ref{Eq:EDO2nL_1}), renders the solution:
\begin{equation}
\frac{-1}{2 \sqrt{-\alpha \psi}} Ln\left( \frac{\alpha \phi - \sqrt{-\alpha \psi}}{\alpha \phi + \sqrt{-\alpha \psi}} \right) = Z + C_1.
\label{Eq:EDO2nL_3}
\end{equation}

Solving for $\phi$ from (\ref{Eq:EDO2nL_3}):
\begin{equation}
\phi = \frac{d u}{d Z} = \frac{\sqrt{-\alpha \psi}}{\alpha} \frac{1 + e^{-2\sqrt{-\alpha \psi}(Z + C_1)}}{1 - e^{-2\sqrt{-\alpha \psi}(Z + C_1)}}.
\label{Eq:EDO2nL_4}
\end{equation}

Under further integration, the solution of (\ref{Eq:EDO2nL_4}) is:
\begin{equation}
u = \frac{\sqrt{-\alpha \psi}}{\alpha} \left[ Z + \frac{1}{\sqrt{-\alpha \psi}} Ln\left( e^{-2\sqrt{-\alpha \psi}Z} - e^{2\sqrt{-\alpha \psi}C_1} \right)  \right] + C_2.
\label{Eq:EDO2nL_5}
\end{equation}

Using the boundary condition in (\ref{Eq:CF}), renders a system of algebraic equations given by:
\begin{equation}
u(Z_1) = \frac{\sqrt{-\alpha \psi}}{\alpha} \left[ Z_1 + \frac{1}{\sqrt{-\alpha \psi}} Ln\left( e^{-2\sqrt{-\alpha \psi}Z_1} - e^{2\sqrt{-\alpha \psi}C_1} \right)  \right] + C_2,
\label{Eq:uZ1}
\end{equation}
\begin{equation}
u(Z_2) = \frac{\sqrt{-\alpha \psi}}{\alpha} \left[ Z_2 + \frac{1}{\sqrt{-\alpha \psi}} Ln\left( e^{-2\sqrt{-\alpha \psi}Z_2} - e^{2\sqrt{-\alpha \psi}C_1} \right)  \right] + C_2.
\label{Eq:uZ2}
\end{equation}

By subtracting (\ref{Eq:uZ1}) from (\ref{Eq:uZ2}), an expression is obtained from which $C_1$ can be calculated, viz.:
\begin{equation}
e^{2\sqrt{-\alpha \psi}C_1} = \frac{e^{-2\sqrt{-\alpha \psi}Z_1} e^{\sqrt{-\alpha \psi}(Z_1-Z_2)} - e^{-2\sqrt{-\alpha \psi}Z_1}}{e^{\sqrt{-\alpha \psi}(Z_1-Z_2)} - 1},
\label{Eq:uC1}
\end{equation}

then,  $C_2$ is:
\begin{equation}
C_2 = \frac{-\sqrt{-\alpha \psi}}{\alpha} \left[ Z_1 + \frac{1}{\sqrt{-\alpha \psi}} Ln \left( e^{-2\sqrt{-\alpha \psi}Z_1} - \right. \right. \\
\left. \left. \frac{e^{-2\sqrt{-\alpha \psi}Z_1} e^{\sqrt{-\alpha \psi}(Z_1-Z_2)} - e^{-2\sqrt{-\alpha \psi}Z_1}}{e^{\sqrt{-\alpha \psi}(Z_1-Z_2)} - 1} \right) \right].
\label{Eq:uC2}
\end{equation}

From these results, the solution for (\ref{Eq:EDO2nL_1}) is:
\begin{equation}
u = \frac{1}{\alpha} Ln \left( \frac{e^{\sqrt{-\alpha \psi}(Z_1(R)+Z_2(R)-Z)}}{e^{\sqrt{-\alpha \psi} Z_1(R)} + e^{\sqrt{-\alpha \psi} Z_2(R)}} \right. + \\
\left. \frac{e^{\sqrt{-\alpha \psi}(Z_1(R)+ Z)} + e^{\sqrt{-\alpha \psi}(Z_2(R)+ Z)}}{\left( e^{\sqrt{-\alpha \psi} Z_1(R)} + e^{\sqrt{-\alpha \psi} Z_2(R)} \right)^2 } \right).
\label{EC:usol}
\end{equation}

\section*{Appendix B. Continuity equation} \label{appB}

To get a useful expression from mass conservation (\ref{Eq:Cont_Adim}), we have the following steps \citep{HoriBOOK2006,RuangStonePRF2024}:

Firstly, (\ref{Eq:Cont_Adim}) is integrated with respect to the axial direction, as:
\begin{equation}
\int_{Z_1(R)}^{Z_2(R)} \left( \frac{1}{R} \frac{\partial}{\partial R}(Ru) + \frac{\partial w}{\partial Z} \right) \; dZ = 0.
\label{Eq:Cont_int_0exp}	
\end{equation}

Under the dimensionless boundary conditions from (\ref{Eq:CF}), i.e., $w(Z_1) = 0$ and $w(Z_2)= - 1$, (\ref{Eq:Cont_int_0exp}) renders as:
\begin{equation}
\frac{1}{R} \int_{Z_1(R)}^{Z_2(R)} \left( \frac{\partial}{\partial R}(Ru) \right) \; dZ - 1 = \frac{1}{R} \frac{\partial}{\partial R} \left( \int_{Z_1(R)}^{Z_2(R)} u \; RdZ \right) - 1 = 0.
\label{Eq:Cont_int_1exp}	
\end{equation}

Integrating (\ref{Eq:Cont_int_1exp}) with respect to the radial direction:
\begin{equation}
\int_{Z_1(R)}^{Z_2(R)} u \; RdZ - \frac{R^2}{2} = 0.
\label{Eq:Cont_int_2exp}	
\end{equation}

Substituting the flow rate (\ref{Eq:flow1}) in (\ref{Eq:Cont_int_2exp}), the continuity equation may be re-cast as:
\begin{equation}
2\int_{Z_1(R)}^{Z_2(R)} u \; RdZ - R^2 = Q - R^2 = 0.
\end{equation}

\end{document}